\documentclass[twocolumn,trackchanges]{aastex631}
\usepackage{lipsum}% for placeholder text
\usepackage{amsmath}

\begin{document}
\makeatletter
\let\frontmatter@title@above=\relax
\makeatother

%Define Commands
\newcommand\lsim{\mathrel{\rlap{\lower4pt\hbox{\hskip1pt$\sim$}}
\raise1pt\hbox{$<$}}}
\newcommand\gsim{\mathrel{\rlap{\lower4pt\hbox{\hskip1pt$\sim$}}
\raise1pt\hbox{$>$}}}

%Colors for edits
\newcommand{\cs}[1]{{\color{burgundy} CS: #1}}
\newcommand{\sn}[1]{{\color{blue} Smadar: #1}}
\newcommand{\kyle}[1]{{\color{magenta} Kyle: #1}}

\newcommand{\GAIA}{\textit{Gaia}}
\newcommand{\COSMIC}{\texttt{COSMIC }}
\newcommand{\POSYDON}{\texttt{POSYDON}}
\newcommand{\norm}[1]{\lvert #1 \rvert}

\title{\Large Triple Evolution Pathways to Black Hole Low-Mass X-ray Binaries: Insights from V404 Cygni
}
\shorttitle{Triple Evolution Pathways to Black Hole Low-Mass X-ray Binaries}
\shortauthors{Shariat et al.}

\author[0000-0003-1247-9349]{Cheyanne Shariat}
\affiliation{Department of Astronomy, California Institute of Technology, 1200 East California Boulevard, Pasadena, CA 91125, USA}
\affiliation{Department of Physics and Astronomy, University of California, Los Angeles, Los Angeles, CA 90095, USA}
\affiliation{Mani L. Bhaumik Institute for Theoretical Physics, University of California, Los Angeles, Los Angeles, CA 90095, USA }

\author[0000-0002-9802-9279]{Smadar Naoz}
\affiliation{Department of Physics and Astronomy, University of California, Los Angeles, Los Angeles, CA 90095, USA}
\affiliation{Mani L. Bhaumik Institute for Theoretical Physics, University of California, Los Angeles, Los Angeles, CA 90095, USA }

\author[0000-0002-6871-1752]{Kareem El-Badry}
\affiliation{Department of Astronomy, California Institute of Technology, 1200 East California Boulevard, Pasadena, CA 91125, USA}

\author[0000-0003-4474-6528]{Kyle Akira Rocha}
\affiliation{Department of Physics \& Astronomy, Northwestern University, 2145 Sheridan Road, Evanston, IL 60208, USA}
\affiliation{Center for Interdisciplinary Exploration and Research in Astrophysics (CIERA), 1800 Sherman, Evanston, IL 60201, USA}

\author[0000-0001-9236-5469]{Vicky Kalogera}
\affiliation{Department of Physics \& Astronomy, Northwestern University, 2145 Sheridan Road, Evanston, IL 60208, USA}
\affiliation{Center for Interdisciplinary Exploration and Research in Astrophysics (CIERA), 1800 Sherman, Evanston, IL 60201, USA}

\author[0000-0001-8220-0548]{Alexander P. Stephan}
\affiliation{Department of Physics and Astronomy, Vanderbilt University, 1221 Stevenson Center Lane, Nashville, TN 37240, USA}

\author[0000-0002-7226-836X]{Kevin B. Burdge}
\affiliation{Department of Physics, Massachusetts Institute of Technology, Cambridge, MA 02139, USA}
\affiliation{Kavli Institute for Astrophysics and Space Research, Massachusetts Institute of Technology,
7 Cambridge, MA 02139, USA}

\author[0000-0002-9751-2664]{Isabel Angelo}
\affiliation{Department of Physics and Astronomy, University of California, Los Angeles, Los Angeles, CA 90095, USA}
\affiliation{Mani L. Bhaumik Institute for Theoretical Physics, University of California, Los Angeles, Los Angeles, CA 90095, USA }

% \keywords{binaries: close, merged – binaries: general – black holes – stars: general, triples – stars: kinematics and dynamics}

\correspondingauthor{Cheyanne Shariat}
\email{cshariat@caltech.edu}

\begin{abstract}

A recent discovery shows that V404 Cygni, a prototypical black hole low-mass X-ray binary (BH-LMXB) is a hierarchical triple: the BH and donor star are orbited by a $1.2$ M$_{\odot}$ tertiary at a distance of at least $3500$ au. Motivated by this system, we evolve a grid of $\sim50,000$ triple star systems, spanning a broad range of initial orbits. Our calculations employ \texttt{MESA} stellar evolution models, using \texttt{POSYDON}, and self-consistently track the effects of eccentric Kozai-Lidov (EKL) oscillations, mass loss, tides, and BH natal kicks. In our simulations, the progenitors of V404 Cygni-like systems have initial outer separations of $1000 - 10000$ au and inner separations of $\sim100$ au, such that they avoid Roche lobe overflow most of the time. Later on, EKL oscillations drive the inner binary to high eccentricities until tides shrink the orbit and mass transfer begins. Notably, such systems only form in simulations with very weak black hole natal kicks ($\lesssim 5\,{\rm km\,s^{-1}}$) because stronger kicks unbind the tertiaries. Our simulations also predict a population of BH-LMXB triples that form via the classical common-envelope channel, when the BH progenitor does overflow its Roche lobe.
The formation rate for this channel is also higher in triples than in isolated binaries because early EKL oscillations cause inner binaries with a wider range of initial separations to enter and survive a common envelope. 
Our calculations demonstrate that at least some stellar BHs form with extremely weak kicks, and that triple evolution is a significant formation channel for BH-LMXBs.  

\end{abstract}

\section{Introduction}\label{sec:introduction}
Black hole low-mass X-ray binaries (BH-LMXBs) are binary star systems where a black hole (BH) accretes material from a low-mass stellar companion ($\lsim1$-$2$~M$_\odot$). The companions are usually main-sequence (MS), subgiant, or giant stars, and their mass transfer creates an accretion disk around the BH \citep{Shakura73}, producing X-ray emission. Currently, about $25$ BH X-ray binary systems have been dynamically confirmed in the Galaxy, most of which are BH-LMXBs \citep{Corral16}. BH-LMXBs mostly reside in quiescent states, characterized by an X-ray luminosity below $\sim10^{32}$~erg~s$^{-1}$, but can exhibit outbursts with X-ray luminosities reaching up to $\sim10^{39}$~erg~s$^{-1}$ for sources accreting near the Eddington limit \citep[see, e.g.,][for a review]{Bahramian23}. BH-LMXBs are widely associated with old stellar populations such as the Galactic center, Galactic bulge, and Galactic clusters \citep[e.g.,][]{Arnason21, Bahramian23}. About half of the known BH-LMXBs are in the Galactic disk \citep[][see figure 9 of the latter for a complete summary]{vanParadijs+95,White96,Grimm02, Tetarenko16,Bahramian23}. Large galactic latitudes are more common among neutron star-LMXBs, likely indicating that they were born with higher kicks \citep{Fragos09,Repetto12,Repetto17,Atri19,Wong14,Kimball23}.

Stellar black holes, including those in BH-LMXBs, are thought to be the remnants of massive progenitor stars \citep[$\gsim20$ -- $40$M$_\odot$, e.g.,][]{Fryer12}. Most stars of this mass do not evolve in isolation and instead have one or more stellar companions \citep[e.g.,][]{Sana12,Kobulnicky14,Moe17,Offner23}. Stellar BH progenitors are theorized to have initial radii above $\sim10$~R$_\odot$, which expand and become significantly larger during the post-main sequence evolution \citep[$\sim1000-3000$~R$_\odot$, e.g.,][]{Levesque05,Romagnolo23}. During this expansion, any companion within $\sim10$~au would likely interact with the primary before it becomes a BH. If the companion's mass is low, as is the case in LMXBs, common envelope (CE) evolution is expected to commence \citep[e.g.,][]{Ivanova+20}, where the low-mass secondary is engulfed in the extended envelope of the BH primary \citep[e.g.,][]{Kalogera+96,Kalogera+98,Kalogera+99,Tauris+06,Taam10,Ivanova13}. 
% Kalogera+96 mostly about kicks, perhaps cite elsewhere

One potential challenge in forming BH-LMXBs through the aforementioned isolated binary evolution channel lies in the common envelope stage. Specifically, considering the energy budget of the system, a low-mass star may not have sufficient orbital energy to unbind the envelope of the black hole progenitor during the unstable CE stage \citep[e.g.,][]{PZ97, Kalogera99, Podsiadlowski03, Justham06}. To resolve this challenge, several ideas were suggested, involving additional sources of energy during the CE phase to successfully eject the envelope \citep[e.g.,][]{Podsiadlowski10,Ivanova11,Ivanova15}. Specifically, it was suggested by \citet{Ivanova11} that the enthalpy of the envelope should be included in the energy budget calculation, which leads to an overall lowering of the binding energy of the envelope. Recent binary stellar evolution studies demonstrated that this channel could then lead to the formation of LMXBs, such as  IC 10 X-1 and MAXI J1305-704 \citep[e.g.,][]{Wong14, Kimball23}. 
Another energy source that may help eject the envelope is nuclear energy sources in the shocked-induced detonation wave that may disrupt the surrounding gas 
\citep[e.g.,][]{Ivanova02, Podsiadlowski10}. Another set of ideas presented in the literature focused on the companion itself. For example, it was suggested that the low-mass companion formed from the disrupted envelope of the massive primary \citep[e.g.,][]{Podsiadlowski95}, or that the companion was initially a larger star and low-mass star observed today is a result of the mass transfer process
\citep[e.g.,][]{Podsiadlowski+00,Chen06,Justham06}. Another set of models relies on dynamically assembling the companion after the BH formed \citep[e.g.,][]{Clark75,Hills76,Voss07, Giesler18, Kremer18}.  

One mechanism that can produce BH-LMXBs without early mass transfer is through three-body dynamics \citep[e.g.,][]{Eggleton86,Michaely2016,NaozLMXB}. A significant fraction of massive stars are born in triple and higher order systems \citep[$68\pm 18\%$,  e.g.,][]{Sana12,Sana14, Moe17, Offner23}. From birth, triple star systems remain stable by naturally tending towards hierarchical configurations \citep{Duchene13}: two stars orbit closely (the inner binary) relative to the tertiary star's wider orbit about the inner binary (the outer binary, see Figure \ref{fig:404_schematic}). 
In hierarchical triples, the inner binary can begin wide enough, avoiding the CE prior to BH formation. Once the primary undergoes core collapse, the binary will then become a detached BH+low-mass main-sequence binary with a separation of $10-1000$~au. At this point, the tertiary can tighten this inner binary through the combined effects of the eccentric Kozai-Lidov Mechanism (EKL), stellar evolution, and tidal capture \citep[e.g.,][]{Kozai1962,Lidov1962,Fabrycky07, Naoz2016,NaozLMXB,Stephan16,Stephan19,Toonen2016,Shariat23,Shariat24,Weldon24}. Secular EKL oscillations from the tertiary can cause extreme eccentricities in the binary that would decrease the periastron of the orbit to less than $10-100$ solar radii \citep[e.g.,][]{NaozLMXB}. At such close periastron distances, tides, magnetic braking, and mass transfer can dissipate orbital energy and angular momentum to shrink the orbit on relatively short timescales, forming a BH-LMXB. Any post-MS stellar evolution of the secondary star increases the probability of earlier tidal locking or mass transfer, making the triple channel even more efficient. This three-body formation scenario has been studied for producing LMXBs in \citet{NaozLMXB}, as well as other accreting compact objects \citep{Toonen2016, Stephan19, Shariat23, Shariat24}. 

One process that may inhibit the triple formation scenario is the presence of a BH natal kick. From stability arguments, the tertiary of a wide BH binary must be at least $5-10$ times further from the inner binary than the separation of the inner binary \citep[e.g.,][]{MA2001}. This means that typical tertiary distances in BH triples are $500$ -- $10^4$~au \citep{NaozLMXB}. At these distances, the tertiary is so weakly bound that even a small kick would unbind the orbit in most cases. The prevalence and typical magnitude of BH natal kicks are uncertain. Theoretical studies have proposed mechanisms where BHs can form through nearly complete implosions with negligible kicks \citep[e.g.,][]{Woosley+95,Sukhbold+16,Mirabel17}.  
Studies of different BH X-ray binaries have derived a wide range of natal kick velocities for different systems. 
Many studies rule out kick velocities greater than $80$-$100$~$\text{km~s}^{-1}$ \citep[e.g.,][]{Mandel16,Nagarajan24} and often support small (to null) natal kicks \citep[e.g.,][]{Reid14, Mirabel17, Shenar22}, with the exception of some systems, such as XTE J1118+40 and MAXI J1305-704, \citep[e.g.,][]{Fragos09, Andrews22, Dashwood24, Kimball23}.

Recently, \citet{Burdge24} discovered that V404 Cygni is orbited by a wide tertiary at $3500$~au separation, making V404 Cygni part of a hierarchical triple system (see Figure \ref{fig:404_schematic} for a schematic).
While the possibility of the existence of tertiary companions was suggested in the past for other systems, \citep[e.g.,][]{Grindlay88,Corbet+94,Thorsett+99,Chou01,Prodan+15,Dage+24}, V404 Cygni is the first robust detection of a BH-LMXB with a wide tertiary companion. At a distance of $2.4$~kpc, the system contains a $\sim9$~M$_\odot$ BH \citep{Khargharia10} accreting from a $0.7$~M$_\odot$ evolved companion with radius $R \sim 6\,R_\odot$ \citep{Shahbaz94} at a separation of $0.14$~au ($P_{\rm orb}=6.4$~days). This makes it one of the widest known BH-LMXBs \citep[e.g.,][]{Corral16}. Additionally, spectral fitting of the tertiary showed that it is beginning to evolve off of the main sequence and is currently at twice its initial radius \citep{Burdge24}. Through isochrone analysis, \citet{Burdge24} constrain the mass of the tertiary to $\sim1.2$~M$_\odot$ and the system's age to $3-5$ billion years. The fact that the K-giant companion in the V404 LMXB is more evolved than the tertiary indicates that the inner companion was initially more massive than $1.2$~M$_\odot$, or equivalently, has since lost at least $0.5$~M$_\odot$ through accretion onto the BH.

Here, we investigate the evolutionary history of V404 Cygni and the broader population of BH-LMXBs with wide companions. We specifically test formation pathways and the impact BH natal kicks have in such systems and compare them to isolated binary formation models. In Section \ref{sec:methodology} we outline the three-body simulations. In Section \ref{sec:orbital_configuration}, we discuss the results from the simulations and their implications for BH natal kicks. We also compare our isolated binary models to the triples to predict the likely formation of Galactic BH-LMXBs. In Section \ref{sec:time_LMXB}, we discuss the time of LMXB formation and how it can potentially distinguish different formation pathways. In Section \ref{sec:spins_inc}, we provide a prediction for the outer tertiary's orientation, in Section \ref{sec:concl} and Section \ref{sec:conclusions} we itemize and discuss our main conclusions, respectively.

\section{Methodology} \label{sec:methodology}

\subsection{Simulations}\label{sec:simulations}
\subsubsection{Three-Body Dynamics}\label{subsec:triple_dynamics}
\begin{figure}
    \includegraphics[width=0.47
    \textwidth]{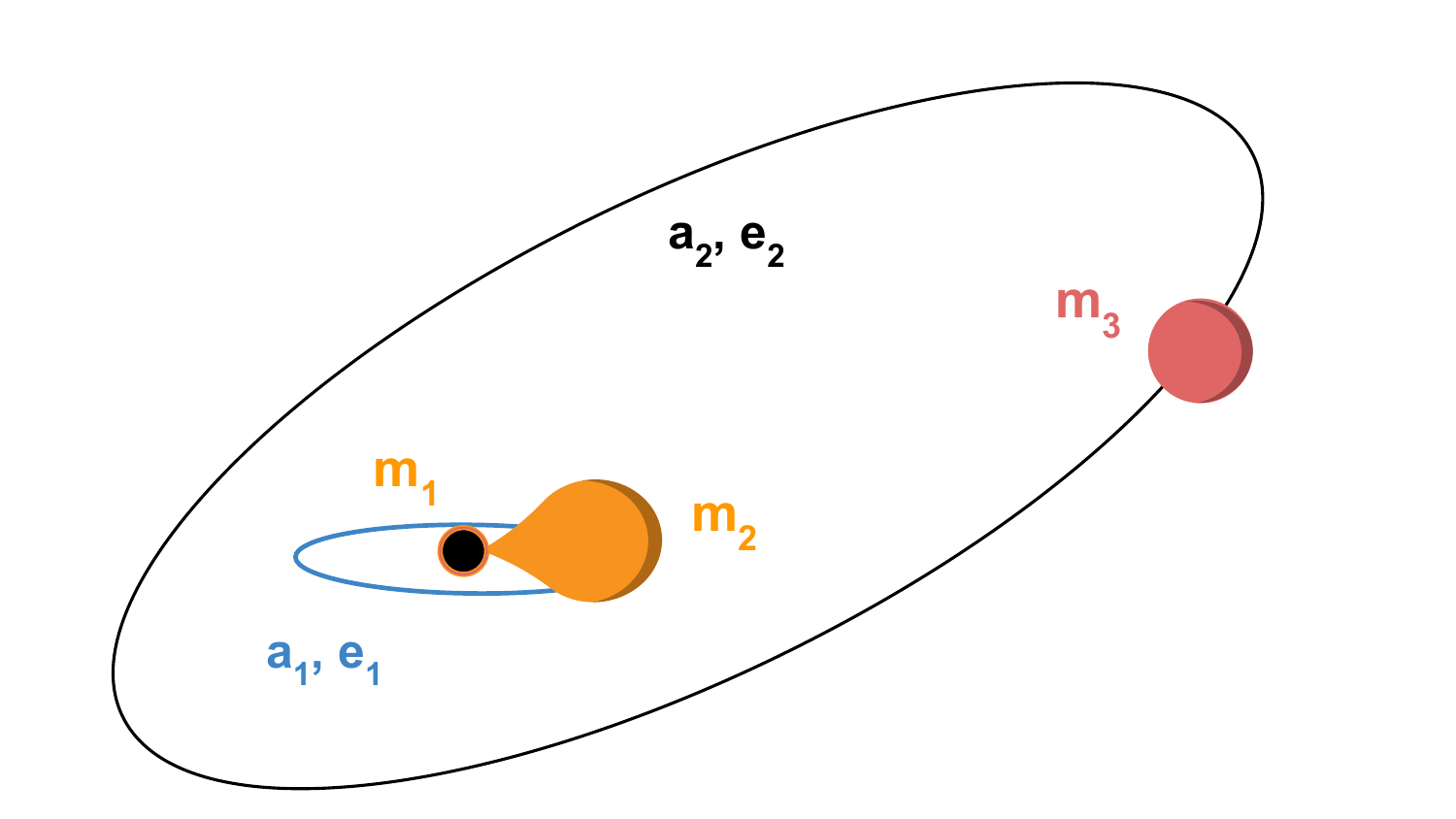}
    \caption{Schematic of V404 Cygni in a hierarchical triple. The triple system contains a close `inner binary' with a semi-major axis and eccentricity of $a_1$ and $e_1$, respectively. The distant tertiary orbits the inner binary, creating the `outer binary', which has a respective semi-major axis and eccentricity of $a_2$ and $e_2$. In the inner binary, a $m_1=9$~M$_\odot$ black hole accretes from a $m_2=0.7$~M$_\odot$ evolved companion in a circular orbit with $a_1\sim0.1$~au, classifying it as a low-mass X-ray binary. At a separation of $3500$~au away, $m_3=1.2$~M$_\odot$ orbits the inner binary, which is also an evolved star. Based on the evolved tertiary, the system's age is $4\pm1$~Gyr. This layout is widely similar for all BH-LMXBs with wide companions. }\label{fig:404_schematic} 
\end{figure}

Throughout this study, we consider a hierarchical triple system with masses $m_1$, $m_2$, in the inner binary, and $m_3$ on a wider orbit about the inner binary (the tertiary). The triple has an inner (outer) semi-major axis $a_1$ ($a_2$), eccentricity $e_1$ ($e_2$), argument of periapsis $\omega_1$ ($\omega_2$), and inclination with respect to the total angular momentum vector $i_1$ ($i_2$). See Figure \ref{fig:404_schematic} for a schematic representation of such a hierarchical triple.

In our simulations, we solve the hierarchical three-body equations of motion up to the octupole level of approximation \citep[see][for the full set of equations]{Naoz2016}. We also include the effects of general relativistic precession for both the inner and outer orbit to 1st post-Newtonian order \citep[e.g.,][]{Naoz2013GR}. In the mass ratios studied here, the first-order description is sufficient to model the dynamics \citep[e.g.,][]{Naoz2013GR,Lim20,Kuntz22}. For the inner binary, we also model the tidal effects by adopting the equilibrium tides model \citep[][see the latter for the full set of the equations in their appendix B]{Hut80, Eggleton98,Kiseleva98,Naoz2016}. This prescription includes tidal precession, rotational precession, and tidal dissipation, the latter of which is modeled assuming a viscous time of $5$~yr \citep[following, e.g.,][]{Naoz2014,NaozLMXB,Stephan16}. Using our tidal model, we can follow the spin precession of stars in the inner binary, which arise from tidal torques and oblateness \citep[e.g.,][]{Naoz2014}. For main-sequence stars with mass greater than $1.5$~M$_\odot$, we use a radiative tidal model \citep{Zahn97}. For red giant stars and low-mass main sequence stars ($<1.5$~M$_\odot$) we assume convective tides \citep{Eggleton98}. The switch between tidal models takes place as a function of stellar type and mass \citep[see][]{Rose+19,Stephan18,Stephan19,Stephan21, Shariat23, Shariat24}. For white dwarfs (WDs), neutron stars (NSs), and black holes (BHs), equilibrium tides are assumed. Magnetic braking is modeled with the stellar evolution part of the code (see below) following  \citet{Rappaport83}.
%Magnetic braking is modeled with {\tt POSYDON}, which follows  \citet{Rappaport83}.

\subsubsection{$\POSYDON$ Single Stellar Evolution}\label{subsubsec:posydon}

In hierarchical triples, the changes in stellar masses and radii associated with stellar evolution can impact the evolution of the system by re-triggering or suppressing EKL \citep[e.g.,][]{Naoz2016, Stephan16}, expanding the inner orbit's semi-major axis faster than that of the outer binary \citep[e.g.,][]{PK12,Shappee13,Michaely2014,Naoz2016}, circularizing the inner binary via tidal interactions \citep[e.g.,][]{Liu15, Bataille18,Angelo22}, leading to mass transfer \citep[e.g.,][]{Salas19,Toonen20,Shariat23}, or even causing a complete stellar merger \citep[e.g.,][]{Antonini16,Toonen18,Stephan16, Stephan19,Shariat24}. However, nearly all previous studies that examine stellar evolution in a triple framework use the rapid fitting formulae of {\tt Single Stellar Evolution} \citep[{\tt SSE},][]{SSE,BSE}, or a variation thereof. The evolution of the component stars can be more accurately followed using detailed stellar models such as MESA \citep{Paxton11}. However, these codes are not often employed for running a large number of simulations because they are slower and more computationally expensive than rapid binary population synthesis codes \citep[e.g.,][]{Paxton19,SSE}. 

Recently, \citet{POSYDON} developed $\POSYDON$: POpulation SYnthesis with Detailed binary-evolution simulatiONs. $\POSYDON$ is a general-purpose code that is capable of evolving a population of binaries on a framework that uses self-consistent stellar evolution models. $\POSYDON$ adopts detailed single binary evolution tracks computed with the Modules for Experiments in Stellar Astrophysics \citep[\texttt{MESA};][]{Paxton11, Paxton13, Paxton15, Paxton18, Paxton19, Jermyn23}. In this analysis, we use {\tt MESA} single-star tracks and interpolation routines distributed with {\tt POSYDON~v1} to follow the evolution of the component stars. Single-star {\tt MESA} models self-consistently model the star's structural response to mass-loss through winds, leading to significant deviations in stellar properties (e.g., final masses and radii) compared to SSE \citep{POSYDON}. 
 
To perform single stellar evolution in $\POSYDON$, we place a single star into a non-interacting binary and only focus on the evolution of the primary star of interest. Specifically, we place our primary star of interest (with given initial mass, metallicity, and spin) into a wide ($a>10^5$~au) circular binary with a $0.5$~M$_\odot$ secondary star. This configuration, a wide orbit with a low-mass secondary, eliminates any chance of the secondary affecting the properties of the primary and is effectively just a single stellar evolution of the primary. Next, we generate a time series of the parameters for the primary for a given evolution time (often $10$~Gyr here). {\tt POSYDON~v1} evolves each star in a detached binary using MESA grids, making our method of single star evolution equivalent to interpolating on single-star {\tt MESA} tracks, with a parameterized CO core mass to compact object mass relation.

{\tt POSYDON} includes several subgrid prescriptions for predicting compact object mass (BH or NS) based on the properties of the progenitor before core collapse. Note that although the different prescriptions share many similarities, they have noticeable differences in the final compact object masses near the NS/BH progenitor mass boundary. Specifically, we use two models. The first is the default core-collapse model from \citet{Patton20}, which utilizes the average carbon abundance at carbon ignition to determine the explodability of the core and assumes that BHs form only from a failed explosion in direct collapse. In this scenario, a $\sim21$~M$_\odot$ progenitor star collapses into a $\sim9$~M$_\odot$ BH. The second core-collapse model we utilize is the \citet{Fryer12} \textit{delayed}. In this channel, a $\sim28$~M$_\odot$ progenitor star collapses into a $\sim9$~M$_\odot$ BH. 
% {\bf For the final BH masses considered in this work, we find that the two models resulted in similar triple population results.} 

We determine the secondary's stellar type based on its stellar structure at different points in its evolution, as derived from the {\tt POSYDON} models. For Figure \ref{fig:periastron_Cyg404_Combined} and Figure \ref{fig:Cyg404_a1_e1_a2}, we bin all stars on the giant branch into the `RG' category to distinguish these evolved stars from those on the MS.

% by assessing the radius of the star. All of the secondary stars in the inner binary began with $R_2=1-1.5$~R$_\odot$, so we consider an `RG' to be any evolved star with $R_2>2.0$~R$_\odot$, though most have radii much larger than this cutoff. {\bf We note that these stars are not strictly red giants, and we instead use this radius cutoff to distinguish evolved stars from those on the MS.} Secondary stars with $R_2<0.02$~R$_\odot$ are considered white dwarfs, and the rest are considered to be on the main sequence. 

Most previous studies of triples used {\tt SSE} or {\tt BSE} \citep{SSE,Hurley02} to model stellar evolution, which can radically alter the evolution of a binary or triple compared to {\tt POSYDON}. One notable difference is that the wind prescriptions in SSE are highly optimistic, and likely outdated, especially for massive stars \citep{SSE, Paxton15, vanSon24}.
For example, a $22$~M$_\odot$ zero-age main-sequence (ZAMS) star in SSE expands to almost $7$~au during its red supergiant phase, which is nearly $3$~au greater than predicted from {\tt POSYDON}, or equivalently {\tt MESA}, at solar metallicity. Though these models are likely more consistent than {\tt SSE}, they are still uncertain \citep{Romagnolo23}.
 Main-sequence stars also evolve more quickly in {\tt SSE}, making it difficult to compare timescales or ages from simulations to observed stellar populations.
Furthermore, in our analysis, we notice that the supergiant phase lasts longer in SSE, and the mass loss associated with it occurs in many discrete steps over the course of Myrs. If the star is embedded in a hierarchical triple, this evolution alters the dynamics and subsequent evolution of the entire triple on these timescales and creates a larger probability of tidal locking. Each discrete mass loss episode changes the mass ratios in the triple, which thereby changes the dynamics during these periods. We touch on this difference more in Section \ref{subsec:example_evolution}. Three-body evolution is generally less sensitive to uncertainties in stellar evolution when the stars are not transferring mass before BH formation. 
All of the results that follow in this work leverage single-star modeling from {\tt MESA} to evolve all three stars in the hierarchical triples.

\subsubsection{Formation Kicks}\label{subsubsec:kicks}
Another new addition to our triple code is the consideration of kicks (with kick velocity $v_k$), including natal kicks during WD, NS, and BH formation. In general, kicks can tilt orbits, change their eccentricities, or unbind them completely. Neutron stars are inferred to experience natal kicks on the order of $100$s of $\text{km~s}^{-1}$ \citep[e.g.,][]{Lyne94,Hansen97, Lorimer97, Cordes98, Fryer99, Hobbs04}. These kicks are essential to the evolution of NSs with companions as they can account for the spin misalignment in pulsar binaries \citep{Lai95, Kalogera96, Kaspi96, Kalogera00}, %Kalogera98
hypervelocity neutron stars \citep{Fragione17a,Fragione17b,Lu19,Zubovas13,Hoang22,Jurado+24}, and perhaps also the eccentricity of wide NS binaries \citep{EB24a,EB24b} for weaker kicks. 
Kicks are also expected to exist during WD formation due to asymmetric mass loss on the AGB, following \citet{EB18,Shariat23,Shariat24,Stephan2024}. 

Lastly, when kicks are selected in the simulations, we consider the effects of BH natal kicks. Currently, the magnitude of BH natal kicks and their frequency is uncertain. Most estimates are made by combining observed spatial velocities with binary evolution models to constrain kick magnitudes in BH X-ray binaries \citep[e.g.,][]{Mirabel01,Mirabel03,Jonker04,Willem05,Fragos09,Repetto12,Wong14,Repetto15,Repetto17,Mandel16}. However, these estimates are often model-dependent, sometimes conflict with one another, and can be accompanied by large uncertainties.
For example, some studies conclude that some BHs form with relatively high kicks, $v_k\gsim80-100$~$\text{km~s}^{-1}$ \citep{Fragos09, Repetto12,Repetto15,Repetto17,Mandel16,Beer02,Kimball23,Mata24, Nagarajan24}. The BH-LMXB V404 Cygni was initially inferred to have experienced a $\sim65$~$\text{km~s}^{-1}$ BH natal kick \citep{Miller-Jones09}, though this is unlikely to be true given the observed 
wide companion's presence. For example, it was estimated that XTE J1118+480 had a natal kick of $v_k\gsim80$~$\text{km~s}^{-1}$ \citep{Fragos09}. Additionally, another system observed via astrometric microlensing, MOA-2011-BLG-191/OGLE-2011-BLG-0462, placed an upper limit of $v_k\lsim100$~$\text{km~s}^{-1}$ \citep{Andrews22}. The recently dynamically confirmed BH-LMXB, Swift J1727.8-1613, was also estimated to form with $v_k>200$~$\text{km~s}^{-1}$, based on its spatial velocity \citep{Mata24}. Similarly, large space velocities were observed in MAXI J1305-704, leading to a constraint of $\gsim70$~$\text{km~s}^{-1}$ on the BH natal kick magnitude. It is also possible that large spatial velocities are the result of dynamical heating from neighboring perturbers, especially for BH-LMXBs, which are generally old \citep{Zhang12,vanParadijs+95}. Some of these systems may have also formed in stellar clusters, where various dynamical processes can cause runaway velocities \citep[e.g.,][]{Poveda67,Rodriguez15}.

On the other hand, several observations rule out the presence of any significant BH natal kicks. Recently, V404 Cygni, a BH-LMXB with a wide tertiary, was reported to have formed with nearly no kick $v_k<5$~$\text{km~s}^{-1}$, on the basis of its bound tertiary in a wide orbit \citep{Burdge24}. Furthermore, for the massive X-ray-faint binary, VFTS 243, a low observed eccentricity suggests that the BH received a negligible natal kick if any \citep{Shenar22,Vigna24}. Studies of Cygnus X-1, a BH X-ray source, show that it likely formed in situ, with a negligible kick \citep{Mirabel03}. Other studies have also suggested small ($0-80$~$\text{km~s}^{-1}$) BH natal kicks \citep[][]{Reid14, Wong12, Mirabel17, Nagarajan24}. 
For a detailed discussion of the broader landscape of BH natal kicks, see \citet{Mirabel17} and \citet{Nagarajan24}. In this study, we test whether BH-LMXBs in hierarchical triples experience BH formation kicks.

When kicks are activated in our models, every compact object formation triggers an added kick velocity to the newly born compact remnant. The kick velocity for any type of compact object (WD, NS, or BH) is sampled from the Maxwell-Boltzmann distribution:
\begin{equation}\label{eq:kicks}
    f(v_k) = \sqrt{\frac{2}{\pi}}\frac{v_k^2}{\sigma^3}\exp{\left(\frac{-v_k^2}{2\sigma^2}\right)} \ .
\end{equation}
For NS kicks, the median value of $v_k$ is $400$~$\text{km~s}^{-1}$ with a standard deviation of $\sigma=265$~$\text{km~s}^{-1}$ \citep{Hansen97, Arzoumanian02,Hobbs04}, though our triple do not produce an NSs. For BHs, we sample from the NS distribution but scale the kick velocity by $1.4$~M$_\odot$/$M_{BH}$, where $M_{BH}$ is the gravitational mass of the BH in solar masses. This effectively assumes that BH and NS kicks have the same linear momentum. For WDs $v_k$ is chosen from the same Maxwellian in Equation \ref{eq:kicks} but with median value $v_k=0.75$~$\text{km~s}^{-1}$ and $\sigma = 0.5$~$\text{km~s}^{-1}$ \citep{EB18b}. Since the resulting orbital parameters are dependent on the orbital orientation of the binary before the kick and on the direction of the kick \citep[e.g.,][]{Lu19}, we randomize both of these at the time of the kick. The post-kick orbital parameters are calculated analytically following \citet{Lu19}\footnote{see also \citet{Hamers2018,Hoang22,Jurado+24}}, whose prescriptions also determine whether the inner and outer orbit remain bound.

\subsection{Numerical Setup}\label{subsec:numerical_setup}

For all of our models, we fix the initial mass of the tertiary to $m_3 = 1.2$~M$_\odot$, based on observations \citep{Burdge24}. Since the secondary is observed to be more evolved than the tertiary, $m_2$ was likely greater than $m_3$ initially \citep{Burdge24}. Following this observation, we sample the secondary mass from a uniform distribution ranging from $1.2 - 2.0$~M$_\odot$. A $2.0$~M$_\odot$ star would evolve off the MS in $\sim 1$~Gyr, which is $2$ standard deviations below the constrained age, hence we choose $2.0$~M$_\odot$ as our upper companion mass limit. Today, V404 Cygni has a black hole with $m_1$ = $9^{+0.2}_{-0.6}$~M$_\odot$ \citep{Khargharia10}. Therefore, we choose an $m_1$ zero-age main-sequence (ZAMS) mass based on the supernova prescription used in $\POSYDON$ for our particular model. In the models where the \texttt{SN\_STEP} engine is the \citet{Patton20} core collapse model, we assume an $m_1 = 21.7$~M$_\odot$ initially, which leads to a $\sim9.2$~M$_\odot$ BH. In the models where we choose the \texttt{SN\_STEP} engine to be the \citet{Fryer12} \textit{delayed} core collapse model, we choose an $m_1 = 27$~M$_\odot$ initially, which also leads to a $\sim9.2$~M$_\odot$ BH. Testing different core-collapse prescriptions is effectively testing how different amounts of mass loss affect the final orbital structure. However, preliminary results showed a minimal difference between the models with different SN prescriptions. Note that the masses are chosen to model potential initial conditions of V404 Cygni, since it is currently the only known BH-LMXB in a triple. The inner binary masses are typical for most BH-LMXBs \citep[][]{Corral16}.

For all of our models, we sample the inner/outer periods from a log-uniform distribution between $0.1 - 10^4$~years. We draw the inner/outer eccentricities from a uniform distribution and draw the inner/outer inclinations from an isotropic distribution (uniform in $\cos i$). The initial spin-orbit angles are also drawn uniformly. After sampling initial conditions for the triple, we require that these parameters satisfy long-term and dynamical stability criteria. The first criterion simply requires that the triple is hierarchical. To test hierarchy, we adopt the hierarchical criterion $\epsilon$, which describes the pre-factor of the octupole level of approximation \citep[e.g.,][]{Naoz13EKL}
\begin{equation}\label{eq:eps_crit}
    \epsilon = \frac{a_1}{a_2}\frac{e_2}{1-e_2^2} < 0.1 \ .
\end{equation} 
To enforce long-term stability, we apply the criteria from \citet{MA2001}:
\begin{equation}\label{eq:MA_stability_crit}
    \frac{a_2}{a_1}>2.8 \left(1+\frac{m_3}{m_1+m_2}\right)^\frac{2}{5} \frac{(1+e_2)^\frac{2}{5}}{(1-e_2)^\frac{6}{5}} \left(1-\frac{0.3i}{180^\circ}\right) \ .
\end{equation}

Deviation from a complete hierarchy does not necessarily mean an instantaneous breakup of the system or an immediate instability \citep{Grishin17,Mushkin20, Bhaskar21, Toonen2022, Zhang23}. However, since V404 Cygni is in a hierarchical triple configuration today, we take the conservative approach and require only stable, hierarchical systems based on the aforementioned criteria.

We run each triple simulation for an upper limit of $10$~Gyr, but systems are stopped earlier if the inner binary (1) crosses the \textrm{Roche} limit or (2) becomes unbound due to kicks. The latter condition is calculated analytically following \citet{Lu19} for the simulations that included kicks. For the first condition, we require that $a_1(1-e_1)>R_{Roche}$ so that there is no immediate Roche crossing in the inner binary. Here, the Roche limit, $R_{Roche}$, is defined by 
\begin{equation}\label{eq:roche}
    R_{Roche,ij} = \frac{R_j}{\mu_{Roche,ji}} \ ,
\end{equation}
where $j\in{1,2}$ represents the two stars in the inner binary and $R_j$ is the radius of the star with mass $m_j$. $\mu_{Roche,ji}$ is the approximate Roche radius given by \citep{Eggleton83Roche}
\begin{equation}\label{eq:roche_mu}
    \mu_{Roche,ji}= \frac{0.49(m_j/m_i)^{2/3}}{0.6(m_j/m_i)^{2/3} + \ln(1+(m_j/m_i)^{1/3})}
\end{equation}

To determine whether the inner binary of a triple system became an LMXB, we filter for systems that meet these criteria:
\begin{enumerate}
    \item The primary star in the inner binary is a BH.
    \item The pericenter of the inner binary is within two times the Roche radius of the secondary star \citep[e.g.,][]{Fabian75}.
    \item The secondary star is on the main sequence or giant branch.
\end{enumerate}
In Table \ref{tab:ICs}, we summarize the different models we used along with their relevant population statistics. 
While we initially considered various supernova models, most of the statistical analysis (Section \ref{subsec:binary_v_triples}) focus on those with a log uniform (LU) initial period distribution and a `PS20' core collapse prescription from \citep{Patton20} for consistency.

%Smadar is here
\begin{deluxetable}{lccccc}
\caption{Simulations Statistics} \label{tab:ICs}
    \tablehead{
    \colhead{} & 
    \colhead{Model} & 
    \colhead{Model} & 
    \colhead{Model} & 
    \colhead{Model} & 
    \colhead{Model} 
    \\
    \colhead{} &
    \colhead{1} &  
    \colhead{2} &
    \colhead{3} &
    \colhead{4} &
    \colhead{5}
    }
    \startdata
    Kicks  & No & Yes & No & No & Yes \\
    Periods & DM91 & DM91 & LU & LU & LU \\
    \texttt{SN\_STEP} & PS20 & PS20 & F12\it{d} & PS20 & F12\it{d} \\
    $N_{\rm Total}$ & 2101 & 795 & 839 & 15594 & 962 \\
    $N_{\rm Roche}$ & 1576 & 688 & 734 & 13719 & 833 \\
    $N_{\rm BH}$ & 525 & 107 & 301 & 2964 & 129 \\
    $N_{\rm LMXB}$ & 48 & 4 & 8 & 375 & 13 \\
    $N_{\rm V404~Cygni}$ & 14 & 1 & 5 & 102 & 6
    \enddata
    \vspace{1ex}
    {Description of the different simulations used in this study. The `Kicks' row specifies whether supernova and WD kicks were assumed. The `Periods' row specifies the initial period distribution that was used. Here, `LU' refers to log-uniform, and DM91 is a reference to the period distribution in \citet{DM91}. The `\texttt{SN\_STEP}' row specifies the supernova engine used: i.e., the mapping between progenitor properties and BH mass. Here, PS20 refers to \citet{Patton20} and F12$\it{d}$ is in reference to the $\it{delayed}$ core-collapse model from \citet{Fryer12}. $N_{\rm Total}$ gives the total number of simulations run for each model. $N_{\rm Roche}$ gives the number of inner binaries that experienced Roche Lobe Overflow with the BH progenitor, i.e., before the BH formed, according to {\tt POSYDON}. $N_{\rm BH}$, $N_{\rm LMXB}$, and $N_{\rm V404~Cygni}$ respectively are the number of systems that formed a black hole in the inner binary without a mass transfer with the BH progenitor, the number that became LMXBs, and the number of LMXBs that had orbital configuration similar to V404 Cygni.}
\end{deluxetable}

\begin{figure*}[ht]
    \centering
    \begin{minipage}[t]{0.49\textwidth}
        \includegraphics[width=\textwidth]{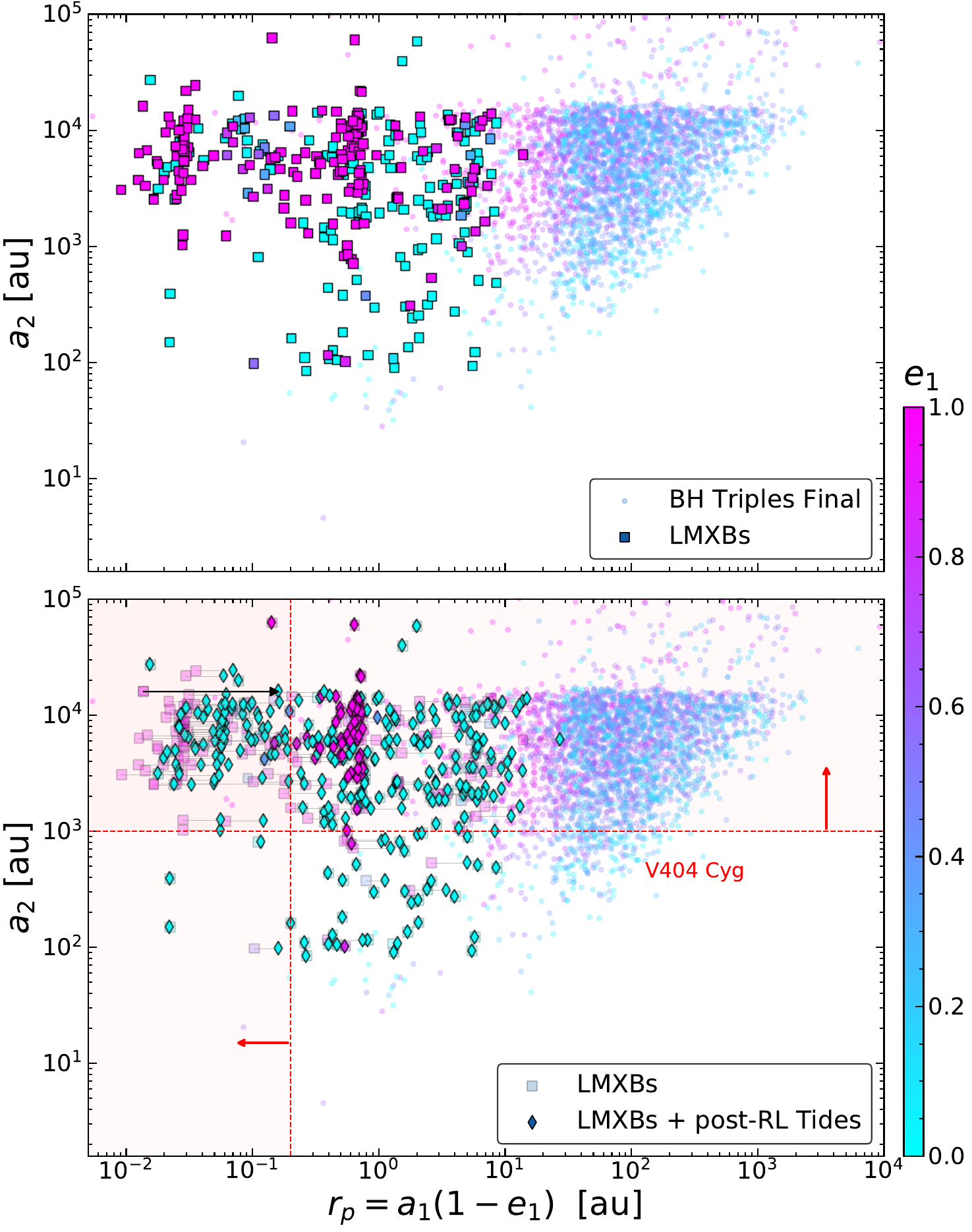}
    \end{minipage}
    \hfill
    \begin{minipage}[t]{0.49\textwidth}
        \includegraphics[width=\textwidth]{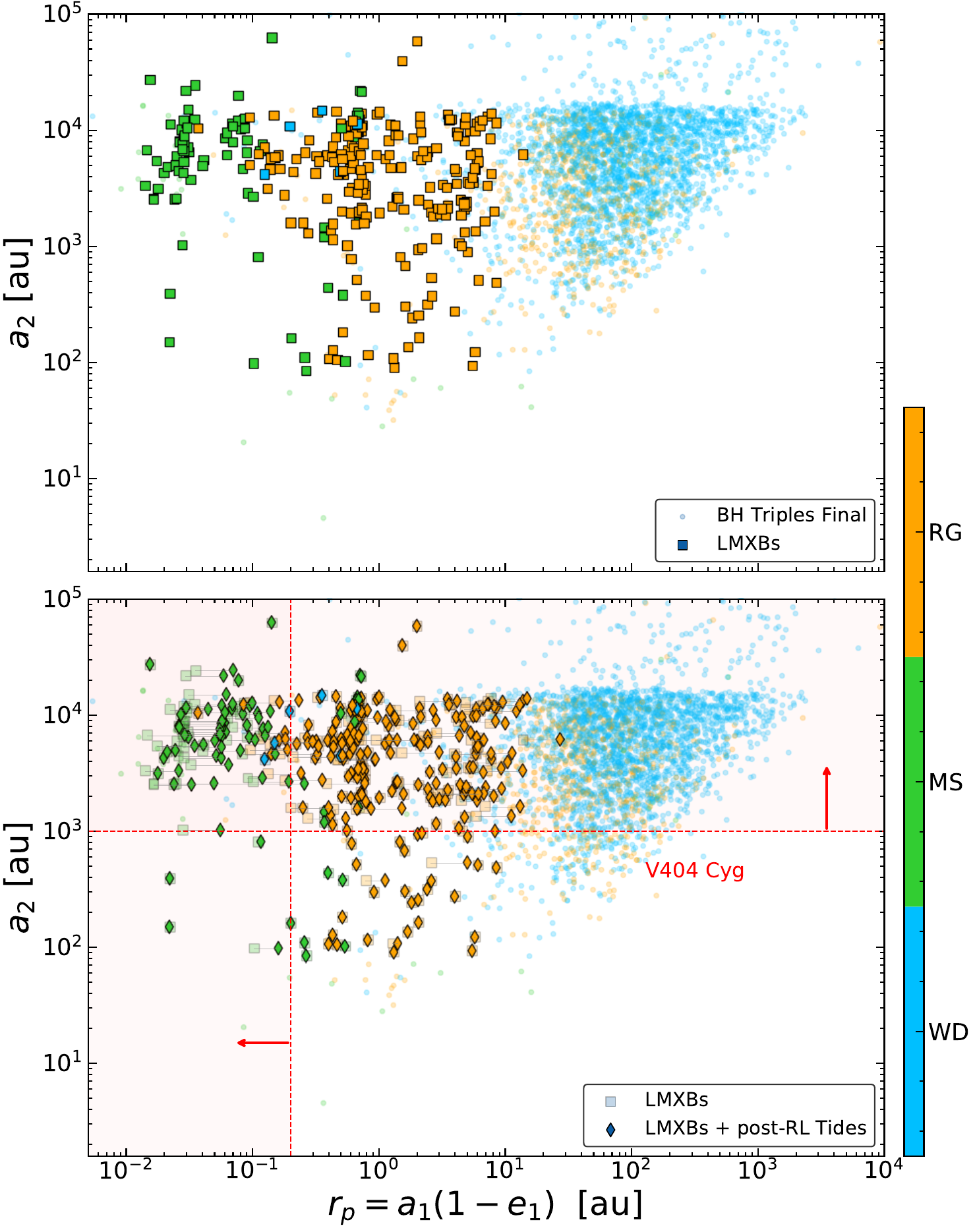}
    \end{minipage}
    \caption{Effect of tides on the periastron of interacting binaries. The top row shows the outer semi-major axis, $a_2$, as a function of the periastron distance at the last step of the simulations.  Square points denote triples with inner BH-LMXBs, and all circular points are detached BH binaries. The bottom row is the same as the top but shows the parameters after tidal evolution. The left and right columns present the same data; only the left is colored by stellar type, and the right is colored by eccentricity. Here, we took all of the LMXBs from the top panel (squares) and evolved their orbits post Roche Lobe crossing as described in Section \ref{subsec:manual_tides}. The post-tides parameters are plotted with diamonds, and a gray line connects the pre-tides to the post-tidal evolution parameters. Tides generally serve to circularize the orbit faster than it is shrunk, resulting in slightly larger periastron distances overall.
    Both columns are identical and only differ by the coloring of the points. The left panel colors by the type of the secondary star, whereas the right panel colors the points by the inner eccentricity, $e_1$.}
    \label{fig:periastron_Cyg404_Combined}
\end{figure*}

% \begin{figure*}[ht]
%     \centering
%     \includegraphics[width=0.99
%     \textwidth]{Figure_7.pdf}
%     \caption{Effect of tides on the periastron of interacting binaries. The top row shows the outer semi-major axis, $a_2$, as a function of the periastron distance at the last step of the simulations.  Square points denote triples with inner BH-LMXBs, and all circular points are detached BH binaries. The bottom row is the same as the top but shows the parameters after tidal evolution. The left and right columns present the same data; only the left is colored by stellar type, and the right is colored by eccentricity. Here, we took all of the LMXBs from the top panel (squares) and evolved their orbits post Roche Lobe crossing as described in Section \ref{subsec:manual_tides}. The post-tides parameters are plotted with diamonds, and a gray line connects the pre-tides to the post-tidal evolution parameters. Tides generally serve to circularize the orbit faster than it is shrunk, resulting in slightly larger periastron distances overall.
%     Both columns are identical and only differ by the coloring of the points. The left panel colors by the type of the secondary star, whereas the right panel colors the points by the inner eccentricity, $e_1$.
%     }\label{fig:periastron_Cyg404_Combined} 
% \end{figure*}

\subsection{Post-Roche Lobe Crossing Tidal Evolution}\label{subsec:manual_tides}

Once the star crosses its Roche Lobe (RL), we expect the inner binary to decouple from the tertiary. We integrate all of these systems forward using our tidal prescription. At this stage we numerically integrate the $a_1$ and $e_1$ evolution \citep[similar to][]{Angelo22} using the coupled tidal equations \citep[][]{Eggleton01}:
\begin{eqnarray}
    \frac{a}{\mid\dot{a}\mid} &=& \frac{t_v f_T(e_1)}{162 (1+2k_s)^2} \frac{m_2^2}{m_1(m_1+m_2)} \left(\frac{a_1}{R_2}\right)^8 \nonumber \\
    & & \times \frac{(1-e_1^2)^{15/2}}{e_1^2}, \label{eq:t_shrink} \\
    \frac{e}{\mid\dot{e}\mid} &=& \frac{t_v f_T(e_1)}{81 (1+2k_s)^2} \frac{m_2^2}{m_1(m_1+m_2)} \left(\frac{a_1}{R_2}\right)^8 \nonumber \\
    & & \times (1-e_1^2)^{13/2} \ . \label{eq:t_circ}
\end{eqnarray}

Here, $R_2$ is the radius of the secondary, $a_1$ is the inner orbit's semi-major axis, $e_1$ is the inner orbit's eccentricity, $m_1$ is the mass of the primary (in our case, the black hole), $m_2$ is the mass of the secondary, $t_v$ is viscous timescale (as mentioned, set to $5$ yr in our simulations), $k_s$ is the classical apsidal motion constant (set to $0.25$), and $f_T(e_1) \equiv 1 + \frac{3}{2}e_1^2 + \frac{1}{8}e_1^4$. Our choice of $k_s$ corresponds to a tidal quality factor of $Q \approx 1.8 \times 10^6$ \citep[see][]{Eggleton01,Naoz2016}. Recall that in the dynamical simulations, we use the complete formalism of \citet{Eggleton01}, which considers spin rates of both stellar components as well \citep{Naoz2016}. Importantly, Equation (\ref{eq:t_shrink}) is the tidal shrinking timescale $t_{shrink} = {a}/{\mid\dot{a}\mid}$, 
and Equation \ref{eq:t_circ} is the tidal circularization timescale $t_{circ} = {e}/{\mid\dot{e}\mid}$. If $a_1=1000 {\rm~au}$, $e_1=0.999$, $m_1=9$~M$_\odot$, $m_2=1$~M$_\odot$, $R_2=1$~R$_\odot$, then $t_{shrink}\sim 1~\text{yr}$ and $t_{circ}\sim 1000~\text{yr}$. These chosen orbital parameters are characteristic of a typical triple in EKL-induced tidal descent and show that tidal forces act on timescales that are generally much quicker than stellar evolution timescales.

Using the Equation (\ref{eq:t_shrink}) and (\ref{eq:t_circ}), we numerically integrate the triples that are halted in the simulations during this phase. We integrate these systems up to $t=5$~Gyr to match the upper constraint on the age of V404 Cygni \citep{Burdge24} and show the results in Figure \ref{fig:periastron_Cyg404_Combined}. Note these timescales evolve nonlinearly in time, as the shorter, shrinking timescale shortens the orbit, allowing for a more rapid circularization. In many cases, tides shrink and circularize the orbits in less than a Gyr. The top row of Figure \ref{fig:periastron_Cyg404_Combined} shows the parameters are the last timestep of the simulations, and the bottom panels include the results of tidal evolution post-RL crossing. In the first column, we color the systems by the stellar type of the secondary. Overall, the smaller, main-sequence secondaries (green points) require shorter pericenter distances to initiate mass transfer or efficient tidal dissipation. On the other hand, the larger red giant secondaries (orange points) become tidally circularized or mass-transferring at larger pericenter separations.

The right panel of Figure \ref{fig:periastron_Cyg404_Combined} colors the points by the inner eccentricity, $e_1$, showing that that about half of the systems halted at higher eccentricities, $e_1>0.9$. In this case, the tidal integration served to circularize and shrink them rapidly (on $\sim100$~Myr timescales). For systems that were already circular (cyan points), the tidal prescription did not change their orbits. For highly eccentric systems, the pericenter actually increases slightly after tides because the system circularizes faster than it shrinks. For all points BH-LMXBs in the figure, high eccentricities ($e_1>0.9$) caused by EKL-induced oscillations from the distant tertiary led to angular momentum loss at close passages and eventual LMXB formation.

\subsection{Example Formation of BH-LMXBs in Triples without Natal Kicks}\label{subsec:example_evolution}

\begin{figure*}[ht]
    \centering
    \includegraphics[width=2.0
    \columnwidth]{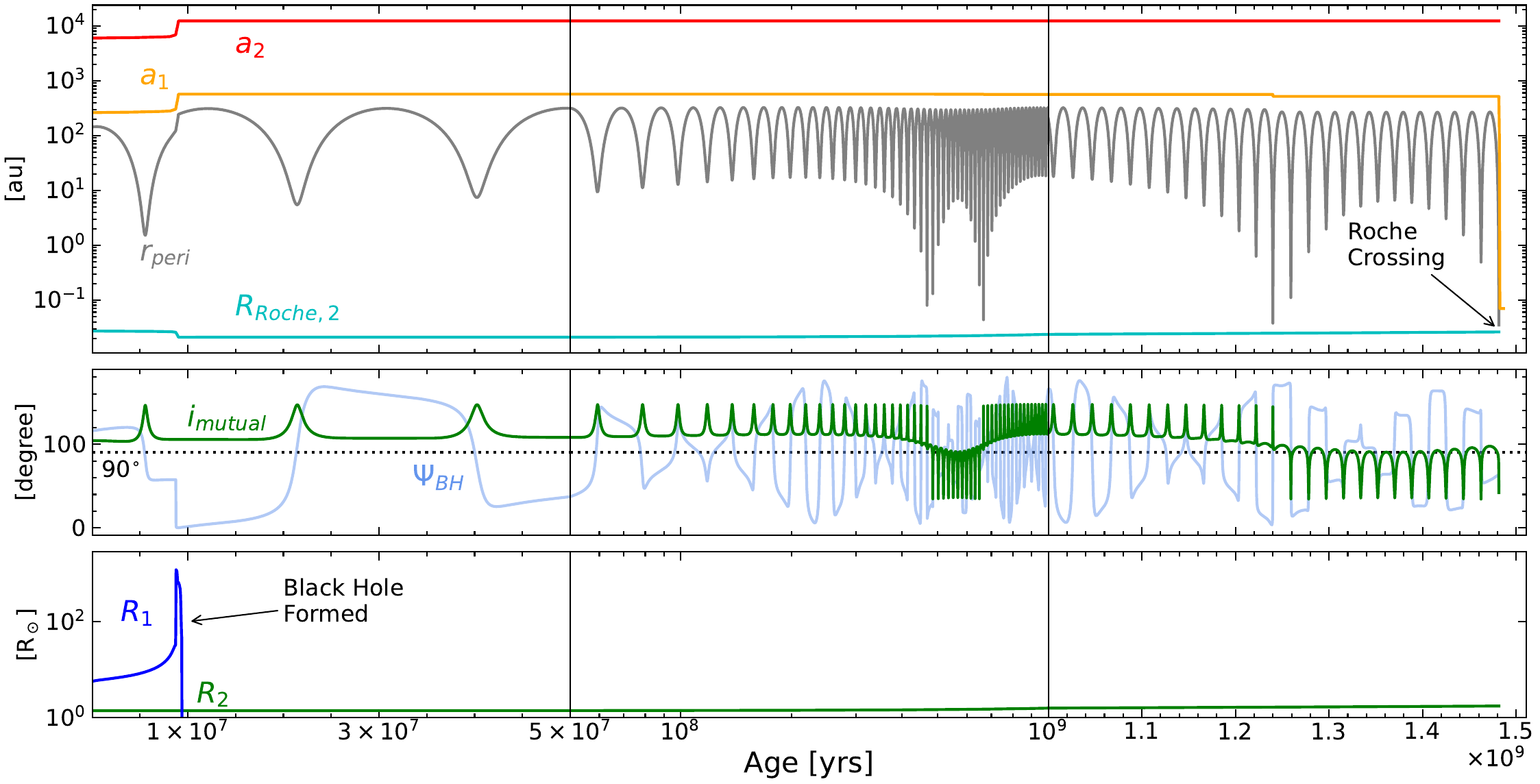}
    \caption{Time evolution of a triple system where the inner binary became an LMXB. {\bf Top:} The evolution of the inner ($a_1$) and outer orbit ($a_2$) semi-major axes (orange and red), the pericenter distance ($r_{\rm peri}$, gray), and the Roche radius of the secondary in the inner binary ($R_{\rm Roche,2}$, cyan). {\bf Middle:} The evolution of the mutual inclination ($i_{\rm mutual}=i_1+i_2$) between the inner and outer orbits (green) and the spin-orbit angle of the black hole($\Psi_{\rm BH}$). Both angles are plotted in degrees. {\bf Bottom:} The evolution of the radius of the primary ($R_1$, blue) and secondary ($R_2$, green) star in the inner binary. We label the moments where the primary star becomes a black hole ($\sim10^7$~yrs) and when the inner binary crosses the Roche limit ($\sim1.5\times10^9$~yrs). In the $500$~Myr leading up to the RL crossing, the secondary star began to evolve off the main sequence with $R_2$ expanding $0.2$~R$_\odot$. Throughout its evolution, the orientation of the inner binary flips multiple times between retrograde and prograde with respect to the outer orbit. At $t=0$ the triples orbital and stellar parameters are $m_1=9.2$~M$_\odot$, $m_2=1.45$~M$_\odot$, $m_3=1.2$~M$_\odot$, $a_1=261.3$~au, $a_2=5976.9$~au, $e_1=0.45$, $e_2=0.75$, $i_1=84.91^{\circ}$, and $i_2=19.25^{\circ}$. At the time of RL crossing, $e_1=0.9999$. Note that the time axes in the middle are log-scaled while the left and right panels are linearly scaled.
    }\label{fig:V404_Timeseries} 
\end{figure*}

In Figure \ref{fig:V404_Timeseries}, we show an example time evolution of a triple where the inner binary became an LMXB under the influence of the tertiary star's effects. In the first $10$~Myr, the primary becomes a black hole. The mass loss from this event expands the orbits of the inner and outer binaries, which undergo secular eccentricity and inclination oscillations through the EKL mechanism. During this time, the inner orbit flips from prograde to retrograde and back multiple times, as shown in the second panel of Figure \ref{fig:V404_Timeseries}.  The flipping, of course, is one of the hallmarks of the EKL mechanism, \citep{Naoz+11}. The eccentricity eventually gets pumped up to $0.9999$, shrinking the pericenter distance to only $\sim6$~R$_\odot$, within the secondary's Roche limit. 
At this stage, the companion had only expanded by $0.2$~R$_\odot$ from ZAMS. 

At the point of RL crossing, the simulation terminates and we evolve the inner binary following Section \ref{subsec:manual_tides}. After only $1$~Myr, the inner orbit tidally shrinks to $a_1=15$~R$_\odot$ and circularizes to create an LMXB triple. Similar to V404 Cygni, this LMXB triple also has a circular inner binary and $a_1\sim 0.1$~au. The final outer orbital elements are $a_2=12278$~au and $e_2=0.77$.  

The particular system that we outline in this section formed from the ``Eccentric" formation channel discussed in \citet{NaozLMXB}. Inner eccentricities above $0.9$ are common for LMXB-forming systems within our sample, and $35\%$ of all LMXBs reached eccentricities above $0.99$. Notably, the triple formation channels that produce LMXBs through eccentricity pumping allow for initially wide inner orbits. Wide orbits that decay after BH formation generally avoid early interactions, which avoids the challenges associated with a CE event.
% In Appendix \ref{app:binary_at_BH}, we discuss the orbits of LMXB progenitors at the time of BH formation.

Since previous triple population synthesis studies used stellar prescriptions similar to those in {\tt SSE} for single stellar evolution, we seek to understand how using {\tt SSE} would compare and affect the outcomes for this system. In Figure \ref{fig:posydon_vs_sse} in appendix \ref{app:posydon_vs_sse}, we evolve the same initial conditions as shown in Figure \ref{fig:V404_Timeseries}, but include {\tt SSE} single stellar evolution in the background (dash-dotted lines). With {\tt SSE}, a $22$~M$_\odot$ ZAMS star only becomes a $\sim4$~M$_\odot$ BH, and the triple does not produce an LMXB. The greater amount of mass loss in the inner binary expands the orbits, weakening EKL. Furthermore, the lower mass ratio in the binary suppresses octupole-level EKL effects, causing lower-amplitude oscillations \citep[e.g.,]{Naoz2016}. Correspondingly, {\tt SSE} evolution causes the system to undergo small eccentricity-inclination oscillations after BH formation and remains detached for its entire evolution. On a population level, the larger stellar radii and smaller mass ratios derived from {\tt SSE} would change the rates of BH-LMXB formation, though the direction of the trend is uncertain. To learn more about the differences between {\tt POSYDON} and {\tt SSE} in the context of triple dynamics, refer to Appendix \ref{app:posydon_vs_sse}. All of the following results leverage {\tt POSYDON}, which calls {\tt MESA} grids, to evolve all stars in the hierarchical triples.

\section{Orbital Configuration of LMXB Triples}\label{sec:orbital_configuration}

\begin{figure*}[h]
    \includegraphics[width=1
    \textwidth]{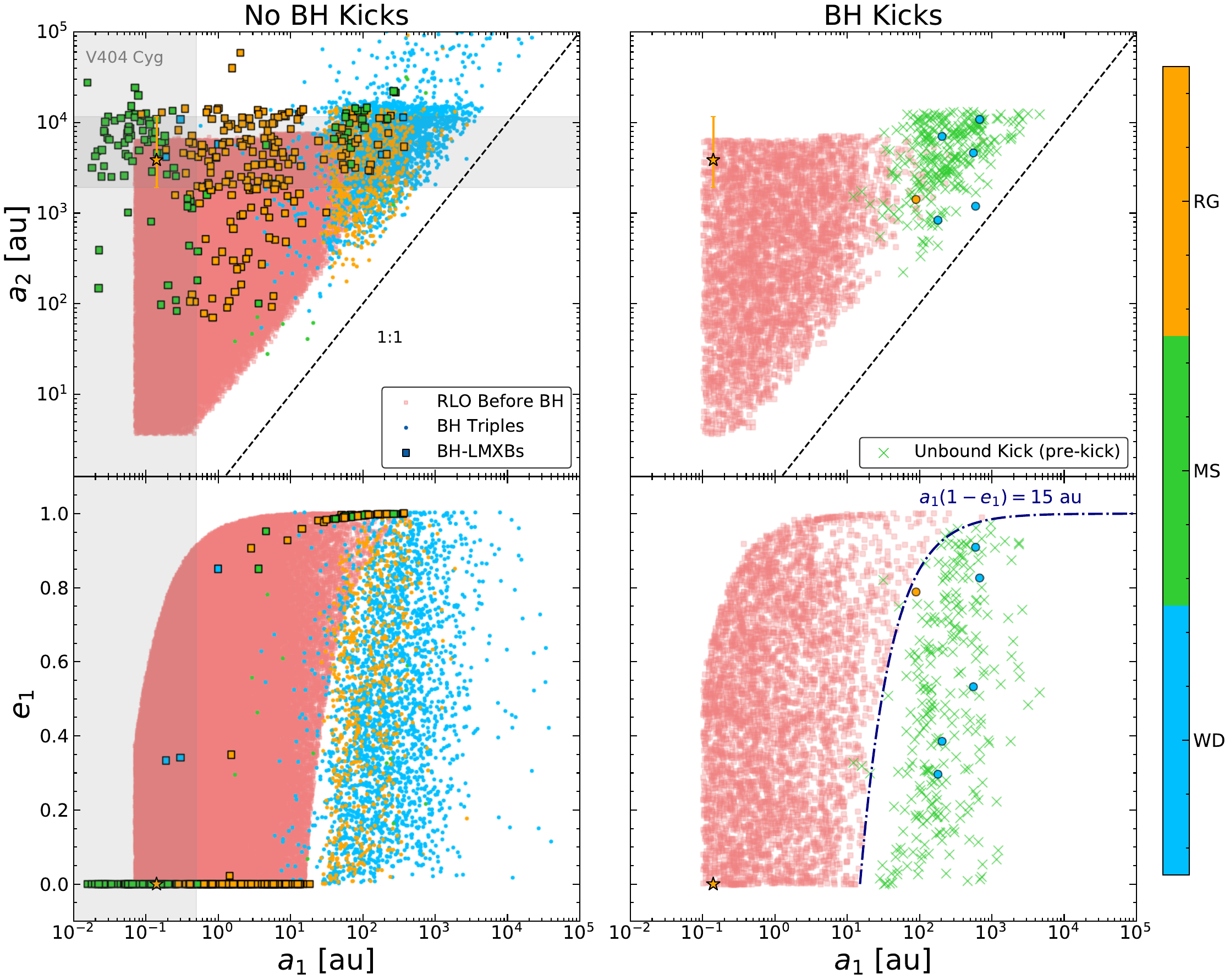}
    \caption{Outer-orbit's semi-major axis ($a_2$) and inner-orbit's eccentricity ($e_1$) as a function of inner-orbit's semi-major axis ($a_1$) at the last simulation timestep. The left column shows the results for simulations that do not include BH natal kicks, while the right column includes BH natal kicks. Red points underwent RL Crossing in the inner binary before the primary BH formed, and most are expected to merge as the result of a failed CE phase; we show their orbital configuration right before RL crossing. The rest of the points have one BH in the inner binary, and the color of these points represents the stellar type of the secondary star in the inner binary (its companion). The secondary stars are either main-sequence (MS, green), red giant (RG, orange), or white dwarf (WD, blue) stars. Circular points are detached BH inner binaries, while the squares are BH-LMXBs: BHs transferring mass with a companion star. The gray-shaded regions are the $a_1$ and $a_2$ observational constraints for V404 Cyg-like systems, which is also denoted by the star scatter point. We only show triples with tertiaries interior to $10^5$~au since those wider than this cutoff would likely be unbound due to the Galactic tide.
    For the systems that became unbound (marked by x's), we plot the orbital configuration just before they became unbound. The kick velocities of the surviving systems range from $2-20$~$\text{km~s}^{-1}$.
    }\label{fig:Cyg404_a1_e1_a2} 
\end{figure*} 

\subsection{Without BH Natal Kicks}\label{subsec:orbital_conf_nokicks}

In Figure \ref{fig:Cyg404_a1_e1_a2}, we plot the orbital configurations for triples with inner BH binaries at the final simulation timestep. The left column displays the simulations that do not include BH natal kicks, whereas the right column of the figure includes kicks. 
The color of each point corresponds to the stellar type of the BH companion in the inner binary. White dwarf, main-sequence, and red giant stars are colored blue, green, and orange, respectively. The square points represent BH-LMXB systems (see Section \ref{subsec:numerical_setup} for LMXB criteria). The small red points denote inner binaries that began RL crossing before the primary BH formed. Since our models do not self-consistently track CE evolution in triples, we simply show their orbital structure at the onset of RL crossing. Note that some of these systems may still become LMXBs, and in later sections, we investigate the outcomes of CE evolution among these inner binaries with {\tt POSYDON}. In the right column, the x's show the orbital parameters of triples that became unbound because of the BH kick. Since they are no longer triples, we plot the orbital parameters just before the kick occurred. The star in the figure plots the observed inner and outer separation of V404 Cygni. We convert the outer separation of $3500$~au into a semi-major axis following the results in Appendix B of \citet{EB18}, which accounts for different underlying eccentricities, inclinations, and observation angles. We conservatively choose the lower (upper) error to be a factor of $1/2$ -- $3$ times the separation. We consider any inner binary with these outer separations and closer than $0.2$~au as `V404 Cygni-like'.
Note that we evolved systems that begin eccentric mass transfer using manual tidal evolution; refer to Section \ref{subsec:manual_tides} and Figure \ref{fig:periastron_Cyg404_Combined} for the post-simulation tidal prescription. Among the systems that formed a BH in the inner binary without early mass transfer with the secondary, $14\%$ become BH-LMXBs, and the rest remain detached.

We find that nearly all of the LMXBs with $a_1<0.2$~au have wide companions at separations of $2,000 - 10,000$~au, just like V404 Cygni. This contrasts with the wider systems ($a_1>0.5$~au) which have a broader distribution of tertiary companions between $10$ -- $10^4$~au. The difference in separations between these two populations effectively distinguishes their distinct evolutionary histories, which can be readily shown in Figure \ref{fig:periastron_Cyg404_Combined}. The LMXBs with $a_1\lsim0.1$~au in Figure \ref{fig:Cyg404_a1_e1_a2} had wider orbits ($a_1\sim100-1000$~au) prior to mass transfer, yet experienced extreme eccentricities ($e_1\gsim0.999$) that made their periastron distances of order $a_1\lsim1$~au (bottom row of Figure \ref{fig:periastron_Cyg404_Combined}). At these distances, tides are dominant and rapidly shrink their orbit to become circular and end at similar separations to their periastron distance during the high eccentricity state (Figure \ref{fig:periastron_Cyg404_Combined}). In most cases, the high eccentricities initiated RL crossing at periastron, often when the secondary star is a red giant (Figure \ref{fig:periastron_Cyg404_Combined}, left column). The high eccentricities required for mass transfer in these systems highlight that their semi-major axes were above $100-1000$~au before the RL crossing. Therefore, their tertiaries are expected to be proportionally 
wider, at separations of $1000-10000$~au from the inner binary. This is precisely where the tertiary separations of LMXBs lie in Figure \ref{fig:Cyg404_a1_e1_a2}, and where the companion of V404 Cygni is today. Refer to Figure \ref{fig:V404_Timeseries} for an example time evolution of such a triple.

For reasons explained above, nearly all BH-LMXBs in our sample with $a_1\lsim0.1$~au harbor companions wider than $10^3$~au, making V404 Cygni's orbital configuration consistent with forming via the triple channel. The separation of V404 Cygni's wide companion, $3500$~au, hints that V404 Cygni likely came from an eccentric channel, where it was previously a wide, detached BH binary that reached extreme eccentricities while the secondary was a giant. At this stage, its orbit decayed significantly through angular momentum loss, either via tidal interactions or mass transfer. In the case of eccentric mass transfer, the systems may radiate in the X-ray, which can classify it as a partial (or micro-) tidal disruption event. Other dynamical channels show that such events can help probe BH populations in dense stellar environments \citep[e.g.,][]{Perets16,Fragione19,Kremer19,Fragione20}.

The slightly wider population of mass-transferring systems, with $a_1\gsim0.5$~au, circularized before mass transfer began. In these triples, the tertiary excites only moderately high eccentricities ($e>0.9$) that do not cause mass transfer during the radial orbit, but tides still circularize and shrink the orbit during close pericenter passages. These tight, circular BH binaries begin to accrete after the secondary star expands, which is shown by the abundance of red-giant secondaries in this region (orange points, Figure \ref{fig:Cyg404_a1_e1_a2}). Stellar evolution of the BH companion plays a strong role in this pathway, which leads most of these systems to become these LMXBs after $\sim1$~Gyr of triple evolution. In Table \ref{tab:LMXBs}, we provide %cite
the number of LMXBs that formed through the different channels and different stellar types of their companions at the onset of mass transfer. 

The separation of LMXBs is generally expected to decay over time as momentum is lost through accretion, outflows, gravitational waves, and magnetic braking \citep[][see latter for a review]{Paczynski67,Verbunt81,Tavani91, Bahramian23}. In our analysis, we do not consider the detailed mass-transfer evolution of the LMXB after its initial Roche crossing, so the separations discussed here will be somewhat different than the current configurations of observed LMXBs. The parameters analyzed here do, however, give insight into the orbital structure of BH-LMXBs before mass transfer and in their early stages. 

\subsection{With BH Natal Kicks}\label{subsec:orbital_conf_withkicks}

To test the impact of natal kicks in BH-LMXBs we run $1642$ Monte Carlo simulations that include BH formation kicks (see Section \ref{subsubsec:kicks} for the kick prescription). We plot the final orbital configuration of these systems in the right column of Figure \ref{fig:Cyg404_a1_e1_a2}. Among all simulations, $84\%$ of the BH progenitors fill their Roche Lobe early on, so those simulations terminate before a kick is induced (small red points). The other $16\%$ ($N=270$) form BHs, and only $6$ of these triples ($2\%$) remain bound after the natal kick. $3$ of the surviving systems had $v_{k} \in 1-7$~$\text{km~s}^{-1}$ and the other $3$ had $v_{k}\sim20$~$\text{km~s}^{-1}$. The kicks for all systems ranged from $0-30$~$\text{km~s}^{-1}$ (see Section \ref{subsubsec:kicks}), and the few that survived had precisely aligned kick angles. Out of the systems that received a kick and remained bound, none became a BH-LMXB, and most of them evolved into detached BHWD inner binaries with moderate eccentricities. From these systems alone, we find a $2\%$ ($6/270$) probability that a wide BH triple remains bound after even a relatively small kick. The $6$ triples that did survive the kicks had similar inner and outer semi-major axes to the no-kick BHs, though the eccentricities and inclinations change. From the simulations without natal kicks, we find $14\%$ of triples that do not transfer mass before BH formation become LMXBs. 
Overall, we find it highly unlikely that any BH-LMXBs with a wide companion, which may be a significant fraction of them \citep{NaozLMXB}, formed with a natal kick. If we take our initial conditions at face value and assume a $5-10$~Gyr total evolution time, we find that for every $10$ detached BH binaries there exist $\sim1$ V404 Cygni-like BH-LMXB (i.e., similar masses and separations). At these rates, the triple channel may be a dominant channel for forming BH-LMXBs (see Discussion in Section \ref{subsubsec:N_LMXBs}). 

\subsection{The Evolutionary Timeline of V404 Cygni}\label{subsec:V404_timeline}

Previously we examined the different formation pathways of BH-LMXBs in hierarchical triples, and here, we discuss the most likely scenario for V404 Cygni specifically. We base this evolutionary history on the observation constraints that the donor is evolved, the age is $4\pm1$~Gyr, and the current inner period is $6.4$~days \citep{Shahbaz94, Burdge24}. Initially, the inner binary began wide ($a_1\sim10-100$~au) allowing for the $20-30$~M$_\odot$ primary to evolve, which widened the orbit even more from mass loss. After $\sim10^7$~yrs, the primary evolved into a BH without a natal kick, forming a wide, detached BH+MS inner binary. Over secular timescales, EKL-induced oscillations from the tertiary excited high eccentricities ($0.9-0.9999$) to the inner binary. After $\sim10^9$~years, the secondary began to evolve off the MS, which allowed tides to shrink and circularize the orbit during one of its radial expeditions (e.g., Figure \ref{fig:V404_Timeseries}). Next, further post-MS evolution caused the secondary to fill its Roche Lobe and begin transferring mass with the BH, creating a BH-LMXB. The subsequent mass transfer and angular momentum loss mechanisms likely further decayed the orbit, creating the BH-LMXB structure that is observed today. The age of V404 Cygni, the evolved state of the donor, and the relatively wide period support this channel.

V404 Cygni could have also formed in through the classical common envelope channel. In this scenario, the role of the tertiary is less clear. If the inner binary was initially tighter than $\sim15$~au, then CE evolution is likely to occur without much dependence of the tertiary's dynamical influence. On the other hand, if the initial separation of the inner binary was slightly larger ($20-100$~au), the tertiary could have excited the eccentricity of the inner binary through EKl oscillations, leading the secondary to fill its Roche Lobe at periastrons. In the latter case, extra orbital energy will be present at the onset of CE, leading to a larger fraction of binaries that survive the CE. However, our models show that both CE channels allow for a wide range of $a_2$ values most of which are inconsistent with the constrained outer separation in V404 Cygni. For EKL-induced CE, the median $\log_{10}(a_2~/~\text{au})$ is $2.27$ with a standard deviation of $0.66$, making this scenario unlikely for V404 Cygni, which has a $\log_{10}(a_2~/~\text{au})$ likely larger than $3.54$ \citep[$3500$~au;][]{Burdge24}. In the regular CE channel, without early EKL oscillations, there is no significant preference for the outer semi-major axis other than potential restrictions from stability. Also, if the rapid non-adiabatic mass loss occurred during the envelope ejection, which is likely \citep[e.g.,][]{Ivanova13}, then a small kick would be imparted on the triple, potentially unbinding the wide tertiary. Given this scenario, and that V404 Cygni's $a_2$ lies in the most probable range for the triple channel without CE (Figure \ref{fig:Cyg404_a1_e1_a2}), we predict that V404 Cygni most likely formed without a CE.
\subsection{Comparison to Binary Evolution Models}\label{subsec:binary_v_triples}
While V404 Cygni, among other BH-LMXBs, likely received no kick, others likely did. In this section, we aim to understand how kicks influence BH-LMXB formation in both triple and isolated binary channels.

\subsubsection{Efficiency of BH-LMXB Formation}\label{subsubsec:lmxb_efficiency}

Using {\tt POSYDON v1} \citep{POSYDON}, we initialize a population of 20000 binaries, 10000 with BH natal kicks and 10000 without.  The initial conditions are identical to the inner binary of our triple population. The primary mass is set to $21.7$~M$_\odot$, which produces a $\sim9$~M$_\odot$ for {\tt SN\_STEP} following \citet{Patton20}. The secondary mass is set to be uniform from $1.2$ -- $2$~M$_\odot$. The initial period is chosen from log-uniform with the period range of $0.1-10^4$~years, and we run all binaries for an upper limit of $10$ Gyr with initially uniform eccentricities and solar metallicities. Similar to the triple simulations, the binary population with BH kicks follows the distribution from \citet{Hansen97} normalized by the black hole mass (see Section \ref{subsubsec:kicks}). Binaries that are merged, disrupted, or unbound due to kicks halt before the $10$~Gyr evolution time. When comparing the results of the binary population to the triples population, we specifically compare the inner binary of the triple to the isolated binary. This allows us to distinguish the role of the tertiary in forming BH-LMXBs.

\begin{deluxetable}{lcccc}
\caption {Outcomes of Triple and Binary Population Synthesis} \label{tab:binary_vs_triple}
    \tablehead{
    \colhead{Outcome} &
    \colhead{Binary} &
    \colhead{Binary+Kicks} &
    \colhead{Triple} &
    \colhead{Triple+Kicks} \\
    \colhead{} &
    \colhead{($\%$)} &
    \colhead{($\%$)} &
    \colhead{($\%$)} &
    \colhead{($\%$)} 
    }
    \startdata
    Merged & $45.2$ & $45.9$ & $87.4$ & $87.6$ \\
    Disrupted  & $1.22$ & $53.0$ & $2.36$ & $10.8$   \\
    Detached BH  & $51.7$ & $0.37$ & $5.49$ & $0.17$  \\
    BH-LMXB & $1.90$ & $0.73$ & $4.58$ & $1.37$    \\
    \enddata
\end{deluxetable}
    % Disrupted  & $1.35$ & $98.26$ & $0$ & $97.78$   \\
    % Detached BH  & $94.66$ & $0.81$ & $85.85$ & $2.22$  \\
    % BH-LMXB & $3.99$ & $0.93$ & $14.15$ & $0$    \\
\begin{figure}
    \includegraphics[width=1
    \columnwidth]{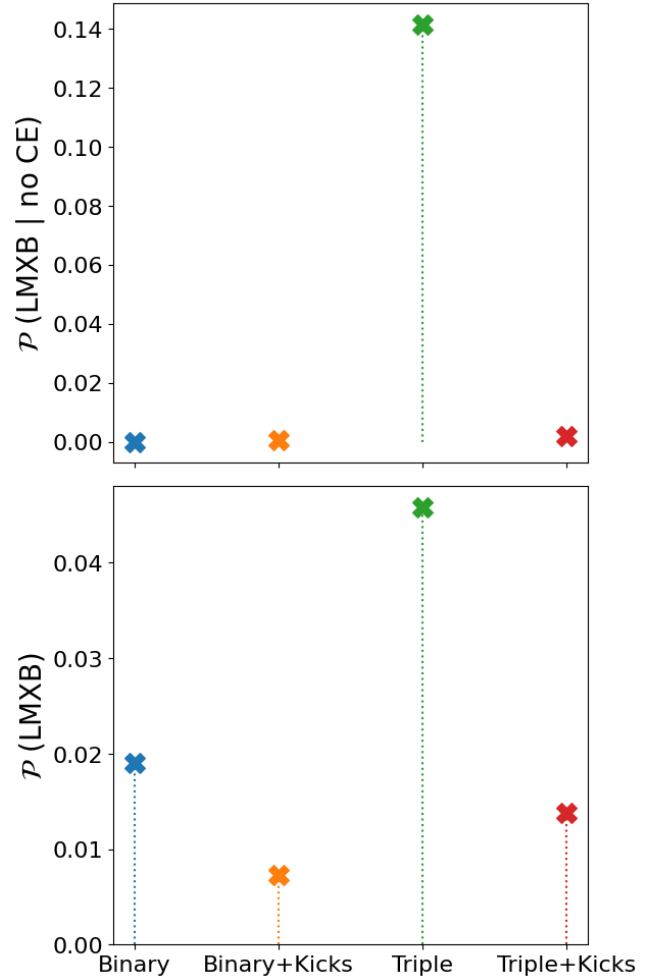}
    \caption{Percentage of BH-LMXBs formed in the different populations. The top row shows the percentage of BH-LMXBs out of all systems that avoided a CE, whereas the bottom row includes all systems in the model's population, including those that underwent a CE phase. Specific values for the bottom panel are listed in the last row of Table \ref{tab:binary_vs_triple}.}
    \label{fig:rates} 
\end{figure}

In Table \ref{tab:binary_vs_triple}, we compare the statistical outcomes from the binary models to the triple models. The four populations include (1) Binaries without BH kicks, (2) Binaries with BH kicks, (3) Triples without BH kicks, and (4) Triples with BH kicks, where the triple populations are the same as analyzed in previous sections (Table \ref{tab:ICs}). For all populations, if the periastron separation of the of BH progenitor with the secondary is less than $\sim10$~au, the stars enter a CE phase. For the binary populations, with and without kicks, $\sim45\%$ experience a CE. In the triple populations, $\sim90\%$ experience a CE. The larger fraction of CE binaries in triples is attributed to the EKL mechanism. Namely, many triples experience early EKL oscillations that excite the inner binary's eccentricity to above $0.9$ before the BH forms. The high eccentricities shrink the periastron distances of many wide inner binaries ($a_1>20$~au) to below the Roche Limit, placing them into a CE phase. In contrast to the binaries, these new CE binaries are wider, supplying more orbital energy for the secondary to eject the envelope of the BH progenitor. The large initial mass ratios ($m_1/m_2\sim20$) make the higher-order EKL effects particularly strong \citep[e.g.,][]{Naoz13EKL}. 

In $\sim95\%$ of cases, the binary does not survive the CE. This fraction is consistent throughout both binary populations since the mass ratios and separations are similar. The remaining $\sim5\%$ that survive the CE end up in tight ($\sim10-100$~R$_\odot$) orbits, most of which become BH-LMXBs. In the kicks populations, many become wide and eccentric from a natal kick, with another fraction unbinding completely. We note that our triples simulations halt when the inner binary begins Roche crossing, meaning that we did not model the CE evolution for the $\sim70-80\%$ of triples that experience a CE. However, since we ran a binary model with identical masses and separations, we identify the closest binary match and determine the mass transfer outcome for the inner binary based on the outcome from {\tt POSYDON}. The outcomes of CE evolution include merging, remaining bound, or becoming unbound due to a natal kick. For the triples+kicks population, we use the CE outcome probabilities from the binaries+kicks populations, and for the triple population, we use the CE outcome probabilities from the binary populations. Assuming similar fractions between the models is not completely accurate because, as mentioned above, the triples more frequently enter a CE at separations wider than the isolated binaries. Therefore, we expect that more triples would survive a CE, given their additional orbital energies. 

The second outcome shown in Table \ref{tab:binary_vs_triple} are the binaries and triples that were made unbound by a natal kick (denoted as `Disrupted'). For the populations with BH kicks, this occurred for $\sim53\%$ of all the binaries and triples. The $1.22\%$ ($2.36\%$) of binaries (triples) that became unbound in the no-kicks population are due to forming neutron stars. Early mass transfer decreased the primary's mass, putting it below the threshold of forming a BH, making it an NS with a natal kick instead. In the binary+kicks population, the binaries that led to BH-LMXBs had kicks of $10-115$~$\text{km~s}^{-1}$ while the detached ones had smaller kicks ($0-50$~$\text{km~s}^{-1}$). For the kick-surviving binaries, the pre-kick separations are bimodal about $0.5$~au and $1.5$~au with circular orbits. After the kick, the bound systems have roughly log-uniform separations between $1$ and $100$~au with moderate to high eccentricities. Overall, since most binaries completely unbind after a kick, the fraction of BH-LMXBs is lower on a population level compared to the no-kick binaries. However, among the minority of bound post-kick binaries, a significant fraction become BH-LMXBs due to their high post-kick eccentricities.

The second category of outcomes in Table \ref{tab:binary_vs_triple} is the `detached BH' which is the fraction of systems within a population that evolves as detached BH binaries, i.e., without ever exchanging mass. In the absence BH natal kicks, $51.7\%$ of binaries remain detached, whereas in triples, only $5.49\%$ remain detached. The lower detached fraction in triples because (1) more inner binaries had a CE, and (2) $14\%$ of detached BH-binaries eventually transfer mass due to high eccentricities caused by the tertiary. When natal kicks are present, the detached fractions are slightly lower than the respective models without kicks because a larger fraction of systems get disrupted, while the ones that survive generally have high eccentricities, leading to a greater mass transfer rate for the surviving systems.

In the last row of Table \ref{tab:binary_vs_triple}, we show the fraction of systems that became BH-LMXBs in each population. Overall, the models without kicks produce a greater fraction of BH-LMXBs. Among all, the triples without kicks are the most efficient, with $4.58\%$ of all systems becoming BH-LMXBs. This large fraction, relative to the binaries and kicks models, is due to the dynamical influence of the tertiary. Firstly, as previously discussed, a tertiary causes more inner binaries to undergo CE evolution at wider separations because it can excite eccentricities while the BH progenitor is evolving, causing close periapsis for binaries that would otherwise stay detached. While a small fraction of these systems survive, those that do are close and often become BH-LMXBs. The second reason for the efficiency of the triple channel is because $\sim14\%$ of wide, detached BH+MS binaries, the tertiary causes high eccentricities that allow tides to shrink the inner orbit and eventually create a BH-LMXB. Therefore, the presence of a tertiary makes BH-LMXB formation $2-3$ times more efficient than isolated binaries ($4.58\%$ compared to $1.90\%$), when BH natal kicks are absent. With BH natal kicks, the triple channel is slightly more efficient because a larger fraction of inner binaries enter a CE, often at wider separations.

Figure \ref{fig:rates} displays the efficiency of LMXB formation for the different channels. The bottom panel reflects the absolute fraction of BH-LMXBs in the population. The top panel shows the same fraction but only considers systems that avoid a CE. The outcome of a CE phase with a mass ratio of $\sim20$ is uncertain, so we show the top panel in the case that all of such CEs result in a merger. In the binary channels, nearly all BH-LMXBs form after a CE event. So, in the pessimistic CE case, the only population that has a significant probability of forming a BH-LMXB while avoiding a CE is the triple without kicks. (Figure \ref{fig:rates})
In the next section, we leverage the rates in Table \ref{tab:binary_vs_triple} and Figure \ref{fig:rates} to compare our theoretical populations to the population of BH-LMXBs in the Galaxy.

\subsubsection{Number of BH-LMXBs in Galaxy}\label{subsubsec:N_LMXBs}

We estimate the number of BH-LMXBs that currently exist in a Milky Way-like galaxy for each of our models, assuming our initial conditions. Considering a star formation rate (SFR) of $1$~M$_\odot$~yr$^{-1}$ and an LMXB lifetime of $\tau_{\text{LMXB}} = 1$~Gyr, we apply the following equation, which is an extension from \citet{NaozLMXB}:
\begin{equation}\label{eq:N_LMXB}
\begin{split}
    N_{\text{BH-LMXB}} = 
    \tau_{\text{LMXB}} \times 
    \text{SFR}\times
    f_{m_1>20}\times  
    f_{q}\times  \\
    f_{\text{binary/triple}}\times 
    f_{\text{BH-LMXB}}
    % f_{\text{bound}}\times 
    % f_{\text{no merge}}\times  
    % f_{\text{BH-LMXB}}.
\end{split}
\end{equation}

For all models, we set $f_{m_1>20} = 2/1000$  and $f_{q} = 1/20$, which are the fraction of stars above $20$~M$_\odot$ assuming a Kroupa IMF \citep{Kroupa2001} and the fraction of systems with a mass ratio of $20$:$1$ assuming a uniform mass ratio distribution \citep{Sana12}, respectively. $f_{\text{binary/triple}}$ is the massive star binary fraction or the massive star triple fraction, depending on whether we are calculating for a binary or triple population. Since the BH progenitor is $\gsim20$~M$_\odot$, we use $f_{\text{binary/triple}} = 0.35$ in the triple models and $f_{\text{binary/triple}} = 0.21$ in the binary models \citep{Sana12,Sana14}. All of the above fractions are independent of the outcomes from our models. The fraction of systems in the population that become BH-LMXBs, $f_{\text{BH-LMXB}}$, is taken from the outcomes of our population synthesis (last row of Table \ref{tab:binary_vs_triple}). Note that $f_{\text{BH-LMXB}}$ is inevitably a function of of our assumed initial masses and separations, making it prone to uncertainties.

Based on these statistics, we find that $N_{\text{BH-LMXB}}=1603$ form in triples and $N_{\text{BH-LMXB}}=400$ form in binaries, if BH natal kicks are not assumed. Combined, they predict a total of $2003$ BH-LMXBs in the Galaxy if BHs are born without natal kicks. With kicks, the triple model predicts $N_{\text{BH-LMXB}}=481$, and the binary model predicts $N_{\text{BH-LMXB}}=153$, for a total of $634$ BH-LMXBs in the Galaxy. If only a fraction of stellar BHs experience natal kicks, then one can adjust these numbers proportionately. For example, if half of the black holes have natal kicks, then $N_{\text{BH-LMXB}} = 1319$ in the Galaxy. Note that both with and without BH natal kicks, we predict that $\sim80\%$ of BH-LMXBs in the Galaxy formed in triple star systems. If kicks are common, these tertiaries are unlikely to be bound today.

By modeling the spatial distributions and outburst recurrence timescales of Galactic BH-LMXBs, \citet{Corral16} estimate that a total of $1280\pm120$ BH-LXMBs exist in the Milky Way. The above estimate assumes a mean outburst recurrence period of $100$~yrs for the transients and relies on the fraction of these systems that have accurate distance measurements. Previously, \citet{White96} and \citet{Romani98} respectively estimate $\sim500$ and $\sim1700$ BH-LMXBs in the Galaxy. All of these predictions also implicitly assume that undetected transients have comparable peak X-ray luminosities to those already observed, so the true number of BH-LMXBs still remains uncertain. Both the kicks and no kicks models are consistent with current estimates for the number of BH-LMXBs. Of course, our population does not encompass the entire landscape of BH-LMXB progenitor masses, and studies with broader initial conditions can improve these estimates.

\section{Time of LMXB Formation}\label{sec:time_LMXB}

As the age of a population increases, the mass of the donors in LMXBs is also expected to decrease \citep{Fragos13a, Fragos13b, Lehmer14}. The age of V404 Cygni is constrained from isochrones to be $4\pm1$~Gyr \citep{Burdge24}, and the present-day BH companion has lost at least $0.5$~M$_\odot$, likely through accretion onto the BH \citep{Burdge24}. Therefore, while the total age of the system is constrained, the precise time that mass transfer commenced, and the LMXB formed, is unclear. In this section, we investigate the time of initial mass transfer in our LMXBs ($t_\text{LMXB}$) and how it relates to various orbital properties.

\begin{figure}
    \includegraphics[width=1
    \columnwidth]{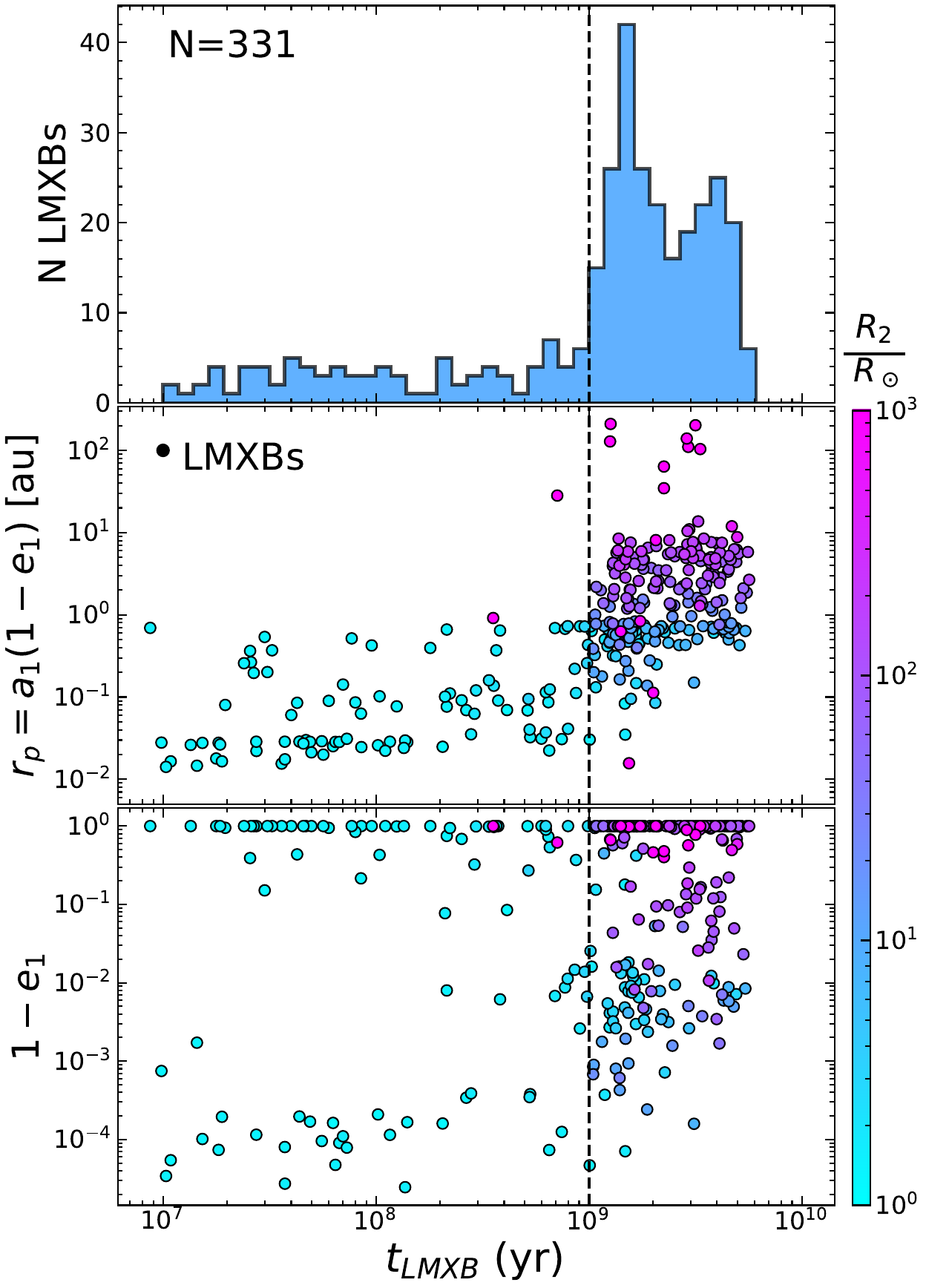}
    \caption{Characteristics of LMXBs at the onset of mass transfer. At the top, we show the distribution of $t_{\text{LMXB}}$. In the middle panel, we plot the pericenter distance, $r_p=a_1(1-e_1)$, at the time of LMXB formation ($t_{\text{LMXB}}$) as a function of $t_{\text{LMXB}}$. In the bottom, we plot $t_{\text{LMXB}}$ vs. $1-e_1$ at $t_{\text{LMXB}}$. The magenta points are red giant (or evolved) BH companions, while the bluer points have MS companions in the LMXB. The black dashed line marks $1$~Gyr. The LMXB population can be divided into the population that formed before/after $1$~Gyr.}
    \label{fig:tLMXB_panel} 
\end{figure}

In Figure \ref{fig:tLMXB_panel}, we display various binary properties against $t_\text{LMXB}$. The top panel shows the distribution of $t_{\text{LMXB}}$, the middle panel shows the pericenter distance ($r_p$) vs $t_{\text{LMXB}}$, and the bottom panel shows $1-e_1$ vs $t_{\text{LMXB}}$. The color of the points corresponds to the radius of the secondary (donor) star in the inner binary. Bluer points mark less evolved, MS stars while the brighter purple points denote evolved giants. Note that the orbital parameters in this figure are at the time that the first mass transfer began between the black hole and its companion, which is often before tides have circularized the orbit. 

Based on the left-skewed distribution, two populations of LMXBs arise: those with $t_{\text{LMXB}}>10^9$~yrs ($N=246$, $73\%$), and those formed before $10^9$~yrs ($N=92$, $27\%$). The panels below the distribution highlight that these different populations correlate to the stellar evolution of the donor star. Nearly all LMXBs with $t_{\text{LMXB}}>10^9$~yrs have companions on the giant branch (Table \ref{tab:LMXBs}). Among the LMXBs with donors on the giant branch, $49\%$ tidally circularized before beginning mass transfer, and $35\%$ began mass transfer while orbit was nearly radial ($e_1>0.9$). Those that circularized formed through a route similar to the ``Giant" channel from \citet{NaozLMXB}, where moderately high eccentricities allowed tides and magnetic braking to shrink and circularize the inner orbit. Then, post-MS stellar expansion of the donor caused Roche Lobe filling and mass transfer, often at separations $\sim1$~au. This channel uniquely requires stellar evolution of the donor star and occurs in $25\%$ of LMXBs. This is a factor of $2$ larger than predicted by \citep{NaozLMXB} and is likely attributed to our focus on masses similar to V404 Cygni. Secondaries with $m_2\gsim2$~M$_\odot$ would evolve in less than a Gyr, while most of those below $m_2\sim1.2$~M$_\odot$  would not become giants in a Hubble time. We also use {\tt POSYDON} \citep{POSYDON}, which has updated stellar evolution prescriptions compared to {\tt SSE} \citep{SSE}, which was used in \citet{NaozLMXB}.

Most of our systems became LMXBs after $1$~Gyr, where most of the secondaries are at least slightly evolved stars (Figure \ref{fig:tLMXB_panel}). This matches closely with observations of V404 Cygni, which show that the secondary is evolved \citep{Burdge24}. In these systems, the pericenter distance is generally larger since a more extended companion will have a larger Roche Lobe. Furthermore, half of the LMXBs with evolved companions tend to be circularized before mass transfer. In contrast, the LMXBs that formed before $1$~Gyr primarily have main-sequence donors. They also possess smaller pericenter distances and more extreme eccentricities, which are required for mass transfer with smaller companions. V404 Cygni has the second most evolved donor among the $\sim25$ BH-LMXBs known. Most BH-LMXBs have MS or slightly evolved MS stars with orbital periods less than $1$~day. Selection effects may also favor short-period systems.

\begin{deluxetable}{lcccccc}
\caption {LMXB Types at Initial Mass Transfer} \label{tab:LMXBs}
    \tablehead{
    \colhead{Formation Channel} &
    \colhead{N} &
    \colhead{$t_{\text{LMXB}}$} &
    \multicolumn{3}{c}{Donor Star Type} \\
    \cline{4-6}
    \colhead{} &
    \colhead{} &
    \colhead{$>\rm1$~Gyr} &    
    \colhead{MS} &  
    \colhead{RG} & 
    \colhead{WD} &    
    }
    \startdata
    Radial~~~($e_1>0.9$)  & $131$ & $91$ & $40$ & $87$ & $4$   \\
    Circular ($e_1<0.05$) & $150$ & $115$ & $35$ & $115$ & $0$  \\
    Other                 & $57$ & $40$ & $14$ & $40$ & $3$    \\
    \enddata
\end{deluxetable}

\section{Spins, Inclinations, and Eccentricities}\label{sec:spins_inc}
\citet{Miller-Jones19} observed that the orientation of V404 Cygni's jets are rapidly changing on the timescale to hours to minutes. It has long been theorized that a tilted disk, which is misaligned from the BH spin-axis, would undergo Lense-Thirring precession \citep{Lense18}. Lense-Thirring precession is a general relativistic consequence of the rotation of a massive body, where the plane of the the orbit precesses about the spin vector of the black hole. For disks, this effect has been tested and confirmed numerically using magnetohydrodynamics simulations \citep{Fragile05,Fragile07}. X-ray outbursts of V404 Cygni are consistent with an optically thick, slim disk configuration \citep{Motta17}, and the inferred mass accretion rate onto the BH implies an outer disk radius consistent for solid-body precession \citep{Miller-Jones19}. Therefore, the precession of the disk in V404 Cygni is consistent with Lense-Thirring precession. We note that these results are somewhat model-dependent and assume intrinsic jets when interpreting the X-ray data. The significance of this assumption, and therefore the results, remains unclear. This supposed misalignment between the BH spin and the binary orbital place of the LMXB was previously attributed to a supernova kick in V404 Cygni \citep{Miller-Jones09, Miller-Jones19} and other BH-X-ray binaries \citep[e.g.,][]{Atri19}. However, from the presence of a distant companion and the kick analysis performed here and in \citet{Burdge24}, V404 Cygni almost certainly formed without a BH natal kick. 

EKL oscillations in hierarchical triples naturally lead to spin-orbit misalignments before the onset of mass transfer\citep{Naoz2014,Liu17,Su21}. Over secular timescales, the inner binary oscillates through a wide range of inclinations, as we illustrate in the second panel of Figure \ref{fig:V404_Timeseries}. When the mass transfer between the BH and its companion begins, the newly-formed LMXB will generally maintain a similar inclination and thereby be misaligned with the BH spin. Here, we examine the angle between the BH spin and the binary orbit, the spin-orbit angle ($\Psi_{BH}$), at the initial formation of the LMXB. 

\begin{figure}
\includegraphics[width=1
\columnwidth]{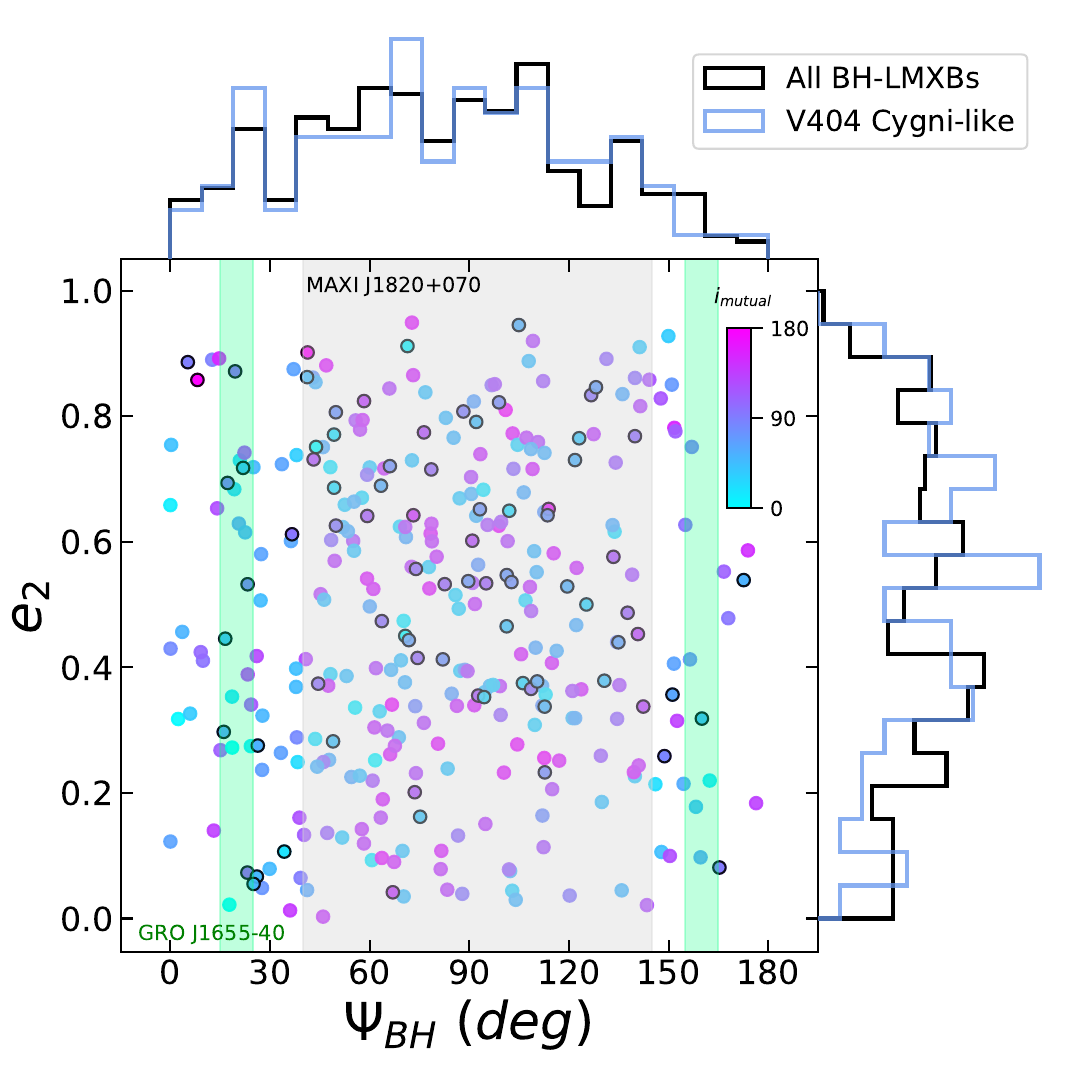}
\caption{Predictions for the eccentricity of the outer orbit ($e_2$) and the spin-orbit angle of the inner black hole ($\Psi_{BH}$). We color the points by the mutual inclination of the triple, which is mostly likely be near $45^{\circ}$ or $135^{\circ}$ (e.g., Figure \ref{fig:404Cig_hist_if}). If an LMXB is outlined in the plot, then it is classified as being V404 Cygni-like, meaning $a_1<0.5$~au and $a_2>2000$~au Interestingly, the $\Psi_{BH}$ distribution is roughly isotropic, implying that most BH-LMXBs formed in hierarchical triples are expected to be misaligned. V404 Cygni, for example, has a likely spin-orbit misalignment, as suggested by the changing orientation of its jets \citep{Miller-Jones19}. The gray and green regions respectively display constraints for the spin-orbit angle of the BH-LMXBs MAXI J1820+070 \citep{Poutanen22MaxiSpin} and GRO J1655-40 \citep{Martin08}.\label{fig:psi_e2}}
\end{figure}

In Figure \ref{fig:psi_e2}, we plot $e_2$, the eccentricity of the outer orbit, against $\Psi_{BH}$, the spin-orbit angle of the black hole (in degrees). The points are colored by the mutual inclination of the triple, and the distributions for $\Psi_{BH}$ and $e_2$ are plotted adjacently. In nearly all cases, triple dynamics will cause a misaligned BH spin-orbit angle in BH-LMXBs. The distribution of $\Psi_{BH}$ is roughly isotropic, with the most likely misalignment angles being between $45^{\circ}$ -- $135^{\circ}$. In the classic formation channel, where LMXBs form in isolated binaries, a primordial BH progenitor undergoes a CE phase with its companion \citep[e.g.,][]{Tauris06}. Up to this event, the BH progenitor expanded significantly, where tidal interactions would have slowed down and synchronized the primary's spin axis with the orbit. Since it is not generally expected that the CE evolution will break the corotation \citep{Ivanova02,Taam10,Ivanova13}. Under this channel, assuming no BH kicks, the BH spin will therefore be aligned with the binary orbit, unlike what is observed in V404 Cygni.

The three-body systems that produce BH-LMXBs slightly disfavor circular outer orbits, $e_2\lsim0.2$, because higher order EKL effects are weaker in this regime \citep[e.g.,][]{Naoz2016}. In fact, we find that LMXB-producing triples mainly had initial outer eccentricities of $e_2\sim0.4$ or $e_2\gsim0.7$ (Figure \ref{fig:404Cig_hist_if}). From the same figure, we find that these systems also had initial mutual inclinations near $90^{\circ}$, meaning the inner and outer orbit were perpendicular at $t=0$. Initially, perpendicular orbits and moderate outer eccentricities both strengthen the amplitude of EKL eccentricity/inclination oscillations in the inner binary of the triple \cite{Naoz13EKL,Naoz2016}. Since the most common LMXB evolutionary pathways in triples include high eccentricities ($\gsim0.95$), such inclinations and eccentricities for the outer orbit are favorable. In its final state, most LMXBs have only moderate mutual inclinations, $|\cos{i_{mutual}}|\sim0.7$, which maps to values of $i_{mutual}$ near $45^{\circ}$ and $135^{\circ}$. 

From the results of our MC simulations that test a wide range of orbital arrangements, we predict that nearly all BH-LMXBs formed through the triple channel will have (1) misaligned BH spin-orbit angles with (2) moderate inclinations and outer eccentricities, irrespective of whether a kick was present or not. We also show that the initial mutual inclination of the triples that form BH-LMXBs is almost always near $90\pm10^{\circ}$. Our prediction for $\Psi_{BH}$ is at the onset of mass transfer, and we do not consider mechanisms that could change $\Psi_{BH}$ during the LMXB evolution. We note that misaligned spins from triples equally include retrograde and prograde spins relative to the binary orbit. Interestingly, \citet{Morningstar14} claim that GS 1124-683, a BH-LMXB, has a retrograde accretion disk.
In contrast to the BH, the donor star is expected to have an exceedingly small spin-orbit angle that has been decayed through mostly tidal evolution \citep[e.g.,][]{Naoz2014, Toonen2016}, which we also find in our simulations. V404 Cygni has already accreted at least $0.5$~M$_\odot$ of material from the companion star \citep{Burdge24}, so any mechanisms that would align initially misaligned orbits would likely have taken place on timescales shorter than its total time of accretion \citep[e.g.,][]{Maccarone02,Martin08,Steiner12,King16}. Beyond V404 Cygni, future observations of a misaligned disk or jet, in the absence of natal kicks, could support the triple formation channel, and perhaps the presence of a hidden companion.

\section{Discussion}\label{sec:conclusions}

As shown here, EKL oscillations enable wide BH inner binaries ($20-1000$~au) to tighten and interact, either by avoiding a CE phase altogether or by increasing the likelihood of successful envelope ejection by initiating the CE at wider separations. Another possibility is that the inner binary in V404 Cygni evolved through a CE phase but with minimal dynamical influence from the tertiary. While this scenario is possible, the observed V404 Cygni's tertiary separation of $3500$~au is among the most probable values predicted by the classical triple channel (Figure \ref{fig:Cyg404_a1_e1_a2}). This supports that the inner binary of V404 Cygni was initially wide and brought together through EKL migration, either before the BH formed -- implying a CE phase -- or after, in which case no CE occurred.

Further, we show that BH-LMXBs that currently reside in triples likely did not receive a BH natal kick or, at most, received a negligible one. Our triple models assume general initial separations but only consider masses similar to V404 Cygni, though these masses are typical for many observed BH-LMXBs \citep{Corral16}. Although dynamically difficult, the presence of a natal kick in BH-LMXB triples is not impossible. Some BH-LMXBs have been constrained to experience kicks of at least tens of $\text{km~s}^{-1}$, often by combining peculiar velocity measurements with galactic orbits and binary evolution models \citep[e.g.,][]{Fragos09,Kimball23,Mata24}. Currently, the distribution of BH natal kicks is not well understood. Some BH-LMXBs seem to have experienced substantial kicks, \citep[$\gsim80~\text{km~s}^{-1}$, e.g.,][]{Fragos09,Repetto12,Repetto15,Repetto17,Mandel16,Kimball23,Mata24} as often inferred from their high peculiar velocities. Others, including V404 Cygni, likely did not receive any significant natal kicks \citep[e.g.,][]{Mirabel03,Wong12,Reid14,Mirabel17,Shenar22,Vigna24,Burdge24}. This discrepancy points to a potential bimodality in the BH kick distribution, where some BHs form with extremely weak to null kicks, while others do not \citep{Nagarajan24}. Here, we provide additional support that some stellar BHs that form without a natal kick. This fraction may be large if a significant number of BH-LMXBs are in hierarchical triples today. The orbits of observed detached BH+luminous companions binaries also support the absence of strong BH natal kicks \citep{EB_BH1,EB_BH2,Chakrabarti23, Wang24}. 

Assuming no BH kicks, we outline the possible separations and orientations of tertiaries to BH-LMXBs (Figure \ref{fig:Cyg404_a1_e1_a2}, \ref{fig:psi_e2}, and \ref{fig:corner_if}). These updated constraints may guide future observations of BH-LMXBs in triples. Nonetheless, detecting BH-LMXB tertiaries remains challenging with current instrumentation. There are $\sim25$ dynamically confirmed BH-LMXBs \citep[e.g.,][]{Tetarenko16,Corral16}, most of which are farther and less optically bright than V404 Cygni. The tertiary in V404 Cygni would not have been detected as a {\it Gaia} proper motion companion if it had been only $\sim1$~kpc farther. As \citet{Burdge24} mention, if the tertiary were only $10\%$ more massive, it would have already been a WD, rendering it too faint for detection. 
If it were $0.5$~M$_\odot$ less massive, then its absolute G-band magnitude would be below the {\it Gaia} detection limit, and if it were only $0.2$~M$_\odot$ less massive, it would not have a precise proper motion measured \citep{Burdge24}. Considering these limitations, we believe that it is highly possible that most BH-LMXBs formed in hierarchical triples, a conclusion that is also supported by our analysis.  

Since LMXBs are associated with older stellar populations \citep[e.g.,][]{Zhang12,Lehmer14} and the tertiaries are generally wide (Figure \ref{fig:Cyg404_a1_e1_a2}), small perturbations on the system may unbind the outer binary. These effects include galactic tides \citep[e.g.,][]{Kaib2014,Grishin22}, stellar flybys \citep[e.g.,][]{Michaely2020}\footnote{Although in some cases these fly-by's can even lead to the formation of BH-LMXBs \citep{Michaely2016}.}, or WD birth kicks, \citep[e.g.,][]{EB18,Shariat23}. Also, if BH-LMXBs in triples experience a CE, rapid non-adiabatic mass loss is likely to occur \citep[e.g.,][]{Ivanova13}, which could unbind wide ($\sim1000$~au) tertiaries.

In this study, we assume that highly eccentric systems that cross their Roche Limit at periastron will rapidly circularize and become BH-LMXBs (e.g., Figure \ref {fig:periastron_Cyg404_Combined}). Currently, the outcomes of eccentric mass transfer are not entirely understood, and some of these binaries could instead manifest as micro-TDEs rather than BH-LMXBs \citep[e.g.][]{Perets16}. Still, a larger fraction of BH-LMXBs in our triple models are already circularized before they begin mass transfer (Table \ref{tab:LMXBs}). Also, though we upgraded our triple dynamics simulations to include new single stellar evolution models, we still do not self-consistently track binary mass transfer within the triple. To estimate the outcomes of an inner binary CE phase, we use {\tt POSYDON}. Compared to older binary evolution codes, {\tt POSYDON's} detailed angular momentum modeling has shown to be a somewhat improved model, especially for Be X-ray binaries \citep[e.g.,][]{Rocha+24}. Incorporating self-consistent binary mass transfer into a triple code would enable the comparison to the broader BH-LMXB population in their current state, though the triple nature of these BH-LMXBs remains unknown.

Our predicted rates for BH-LMXBs in the Galaxy are highly dependent on the results of CE evolution from {\tt POSYDON}, which predict that $\sim92\%$ of the CE binaries will merge. At the separations and mass ratios considered here, it may be the case that essentially all CEs lead to a merger when the mass ratio is $20$.
In this scenario, the triple channel without kicks would continue to produce BH-LMXBs efficiently, while the binary and kick models would not (top panel of Figure \ref{fig:rates}). Future investigations are required to constrain the outcomes of CE evolution and shed light on the formation pathways of LMXBs.

\section{Conclusions}\label{sec:concl}
The recent detection of a wide companion orbiting V404 Cygni has allowed for a unique opportunity to test the formation and evolution of BH-LMXBs. Specifically, the presence of the tertiary requires the BH to have formed with a very weak kick and may support three-body formation channels for LMXBs. Here, we test both of these hypotheses by evolving a large grid of dynamical three-body systems and comparing them to a binary population from {\tt POSYDON}. We revise our triple dynamics code by incorporating single stellar evolution {\tt MESA} tracks, using {\tt POSYDON}. Compared to older codes, these models use improved
prescriptions that include full stellar structure modeling and are consistent with {\tt MESA} stellar evolution, \citep{POSYDON}. This treatment yields more realistic radii for evolved massive stars compared to {\tt SSE}, which alters the evolution and outcomes of triple star systems (see Figure \ref{fig:posydon_vs_sse}). Although we focus on V404 Cygni as a motivating case study, our analysis considers general orbital parameters (e.g., periods, eccentricities), and the initial masses chosen are similar to many other BH-LMXBs \citep{Corral16}, allowing us to apply our results to a broader BH-LMXB population. Our main conclusions are summarized as follows:

% {\bf initial} stellar and orbital parameters considered here are quite general and may be applied to a broader BH-LMXB population. Our main conclusions are summarized as follows:
\begin{enumerate}
    \item {\it Outer separation distribution of BH-LMXBs in triples:} We show that when BH-LMXBs form through the classical triple channel, without a CE phase, the tertiary will have typical separations between $10^3 - 10^4$~au (Figure \ref{fig:404Cig_hist_if}). If a CE occurred, then the tertiary's separation is less constrained, but generally is $\sim100-1000$~au (see Section \ref{sec:orbital_configuration}).
 
    \item {\it BH-LMXBs in triples likely form without black hole natal kicks:} Our analysis shows that BH-LMXBs in hierarchical triples likely experienced no BH natal kicks, or at most, weak ones (with $v_{\rm k,max}\leq 5$~$\text{km~s}^{-1}$; see Figure \ref{fig:Cyg404_a1_e1_a2}). For V404 Cygni, we derive a $<0.2\%$ likelihood that the BH formed with even a small natal kick ($\lsim30$~$\text{km~s}^{-1}$).
   
    \item {\it The triple channel produces BH-LMXBs more efficiently than isolated binaries:} By comparing our population of triples to a binary population from {\tt POSYDON}, we find that triple evolution is the dominant formation pathway for BH-LMXBs. In triples, the fraction of LMXBs formed increases by a factor $2.5$ ($2$) without (with) natal kicks, assuming an equal number of binaries and triples (Table \ref{tab:binary_vs_triple}). However, assuming a triple fraction of $\sim 35\%$ \citep[e.g.,][]{Sana12,Sana14}, a galactic stellar population may yield that $80\%$ of BH-LMXBs form in triples (see Figure \ref{fig:rates} and Section \ref{subsubsec:N_LMXBs}.) 
    While this estimate relies on our assumed initial masses and separations, it applies irrespective of whether BH kicks occur in the system. The boosted rates of LMXB production in triples are attributed to two factors. First, early EKL oscillations spark high eccentricities, leading twice as many
    binaries to cross their Roche Limit and enter a common envelope with the BH progenitor, when compared to isolated binaries. In triples, these Roche crossings typically occur at wider separations with larger orbital energies. Second, when the tertiary remains bound, it can induce secular EKL torques to the inner BH binary, which brings together $14\%$ of wide, detached BH binaries (e.g., Figure \ref{fig:V404_Timeseries}). Note that even with strong kicks, this first mechanism dominates, but the tertiary would have since been unbound from the system. In the pessimistic case, where nearly all CE phases with the BH progenitor lead to a merger, then the no-kicks triple scenario is the only significant contributor to LMXB formation among the channels considered here (top panel of Figure \ref{fig:rates}). Even in this regime, the wide tertiary may become unbound following galactic tides and fly-by interactions with field stars \citep[e.g.,][]{Kaib2014,Grishin22,Michaely2020}.
   
    \item {\it The orbit and orientation of BH-LMXB tertiaries:} We constrain the possible tertiary orbits at birth and at BH-LMXB formation (Figure \ref{fig:404Cig_hist_if}). The triple formation channel predicts that the current-day BH-LMXBs likely formed after $1$~Gyr (Figure \ref{fig:tLMXB_panel}, and their tertiaries most probably have wide separations ($1000$ -- $10,000$~au) with moderate mutual inclinations ($40^{\circ}$ or $140^{\circ}$) and eccentricities ($0.3$ -- $0.8$). 
    Additionally, the spin of the BH in BH-LMXBs, if formed in a triple, will generally be misaligned with the binary orbit at the onset of mass transfer (Figure \ref{fig:psi_e2}). If the spin does not align quickly, the misalignment may manifest observationally as a retrograde accretion disk \citep[e.g.,][]{Morningstar14} or as rapidly changing jet orientations, which was claimed in V404 Cygni \citep{Miller-Jones19}.
    If we consider inner binaries that underwent a common envelope, then the outer orbit is less constrained.

\end{enumerate}

\section{Acknowledgements}
We thank the anonymous referee for their constructive report and Pranav Nagarajan for valuable discussions on black hole natal kicks. C. S. and K. E. were supported in part by NSF grant AST-2307232. S.N. acknowledges the partial
support from NASA ATP 80NSSC20K0505 and from the NSF-AST 2206428 grant, as well as thanks Howard and Astrid Preston for their generous support. This work used computational and storage services associated with the Hoffman2 Cluster, which is operated by the UCLA Office of Advanced Research Computing’s Research Technology Group, and the Resnick High-Performance Computing Center, a facility supported by the Resnick Sustainability Institute at the California Institute of Technology. 
KAR is supported by the Gordon and Betty Moore Foundation (PI Kalogera, grant awards GBMF8477 and GBMF12341) and the NASA grant awarded to the Illinois/NASA Space Grant Consortium, and any opinions, findings, conclusions, or recommendations expressed in this material are those of the author and do not necessarily reflect the views of NASA.

\appendix 

\section{{\tt POSYDON} vs {\tt SSE} Single Stellar Evolution}\label{app:posydon_vs_sse}
\begin{figure}[h]
\centering
\includegraphics[width=1.0
\columnwidth]{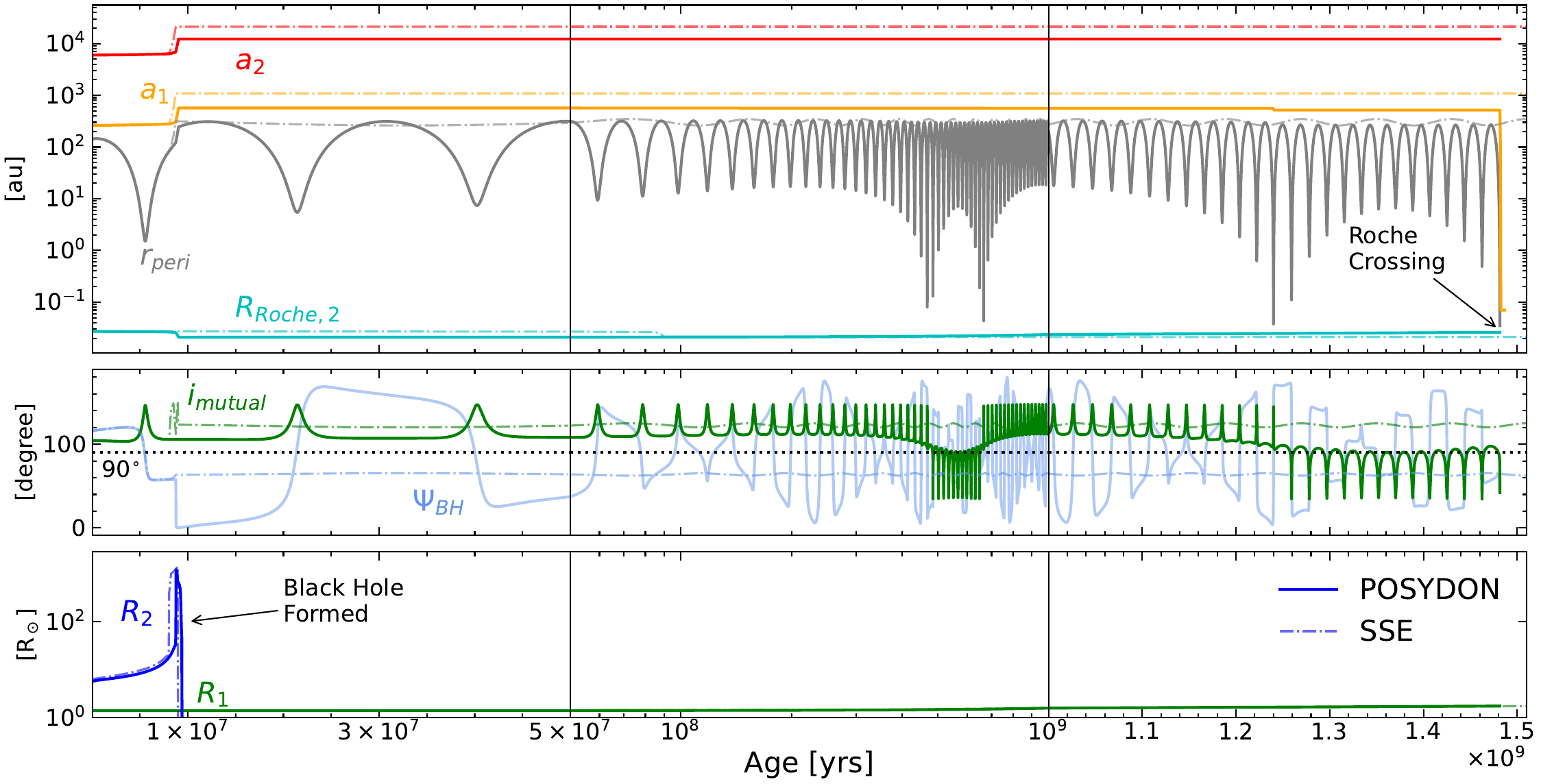}
\caption{{\tt POSYDON} vs {\tt SSE} single stellar evolution as implemented in the complete triples dynamics code. This figure is the same system as Figure \ref{fig:V404_Timeseries}, but the dashed points show the same system evolved using {\tt SSE}. 
}\label{fig:posydon_vs_sse} 
\end{figure}

In Section \ref{subsubsec:posydon}, we discuss our use of  {\tt POSYDON} for modeling the single stellar evolution for all three stars in the triples. In summary, {\tt POSYDON} is a binary evolution code that leverages extensive grids of single- and binary-star models based on {\tt MESA}. The code also implements state-of-the-art interpolation and postprocessing methods, which self-consistently evolve single and binary systems \citep{POSYDON}. The major advance of {\tt POSYDON} for single stellar evolution, as opposed to older codes such as {\tt SSE} or {\tt COSMIC}, is their accurate treatment of massive stellar evolution: i.e., stars that become black holes or neutron stars. Some key changes include updated recipes for stellar winds and core-collapse supernovae, both of which have modular aspects in the code. For the analysis in this paper, we have one massive star in the triple at ZAMS -- the primary BH progenitor -- so we explore the morphological differences between using {\tt POSYDON} vs {\tt SSE}. In Figure \ref{fig:posydon_vs_sse}, we plot the time evolution for the same initial conditions as Figure \ref{fig:V404_Timeseries}, but model single stellar evolution using {\tt SSE} (dash-dotted lines). 

With {\tt SSE}, the three-body evolution completely changed to create a widely different outcome than the original evolution. The main difference comes from the mass loss effects since the $\sim22$~M$_\odot$ primary only produces a $4$~M$_\odot$ BH. This resulted in a weakened impact of the tertiary on the inner binary, which causes only modest EKL oscillations over Gyrs. Furthermore, the result of this evolution was a widely detached ($1000$~au) BH binary, unlike the mass transferring BH binary produced with {\tt POSYDON}+triple evolution. To learn more about the detailed, self-consistent stellar models in {\tt POSYDON}, refer to \citep{POSYDON}.

\section{Tidal Evolution of the inner binary Before LMXB Formation}\label{app:tides}
As discussed in Section \ref{subsec:manual_tides}, some of the simulations get computationally expensive when an evolved secondary star is on a highly eccentric orbit in the inner binary. Over half the time, the internal tidal prescription in the code evolves and circularizes these binaries rapidly. However, in a fraction of systems, the simulations slow down significantly at this stage. On timescales shorter than stellar evolution, tides will often dissipate orbital energy, which will shrink the inner orbit and make it circular. In highly eccentric orbits, as are present here, tides are especially efficient during the close pericenter passages. We, therefore, manually evolve the inner binary using the tidal equations above. Since no mass loss is involved, the tertiary separations remain the same throughout this process.

In Figure \ref{fig:periastron_Cyg404_Combined}, we show the impact of the manual tidal evolution in altering the inner binary configuration for systems that became mass transfer LMXBs. The pericenter distance of the inner orbit ($r_p = a_1(1-e_1)$) is plotted on the horizontal axis while the semi-major axis of the outer orbit ($a_2$) is plotted along the vertical axis. The top panel shows the initial-final periastron distribution. The second panel shows the periastron of the triples at the stop time of the triple code, where the color denotes the inner eccentricity ($e_1$) and the squares denote LMXB-forming triples. The bottom panel is the same as the second, but after applying manual tidal evolution. In the bottom panel, we also plot arrows denoting the direction of evolution for each of the LMXBs. In general, since the tidal evolution circularizes the orbits, the periastron distance for already-circular orbits does not change. For the bright pink systems, which have high eccentricities ($0.9-0.9999$), tides decrease their semi-major axis ($a_1$) while also taking $e_1$ to $0$. The balance between the decreasing ($a_1$) and the increasing $1 - e_1$ generally results in an inner binary orbit that is slightly larger than the initial periastron distance at high eccentricities (bottom panel). One example is shown with the black arrow in the bottom panel, where a triple with $1 - e_1 = 10^{-6}$ and $a_1 = 13621.72$~au tidally decayed over a few periastron passages into $1 - e_1 = 1$ $a_1=0.16$~au. As shown in this example, highly eccentric binaries that were stopped in the code circularize quickly ($e_1\lsim0.01$), which makes the pericenter effectively equivalent to $a_1$. Although the tidal evolution generally occurs on $1$ -- $100$~Myr timescales, which is often shorter than stellar evolution timescales, there is a possibility that the secondary star will expand significantly during the tidal evolution. In this case, an eccentric mass transfer would occur, which can change the orbital structure \citep[e.g.,][]{Sepinsky07b,Dosopoulou16b,Hamers19}, especially within the context of triple dynamics \citep[e.g.,][]{Toonen2016,Toonen18,Hamers19}. We do not follow the detailed mass transfer physics, though mass transfer could likely help remove angular momentum more rapidly \citep{Hamers19}. This would not only create the an LMXB, or Symbiotic binary, at an earlier time but can further help shrink and circularize the orbit. Note also that an expanding stellar radius makes tides exponentially more efficient, which would further compound angular momentum loss (Equation \ref{eq:t_shrink}, \ref{eq:t_circ}).

\section{Initial-Final Relations for BH-LMXBs with wide companions}\label{app:big_plots}
\begin{figure}[h]
    \centering
    \includegraphics[width=0.96
    \textwidth]{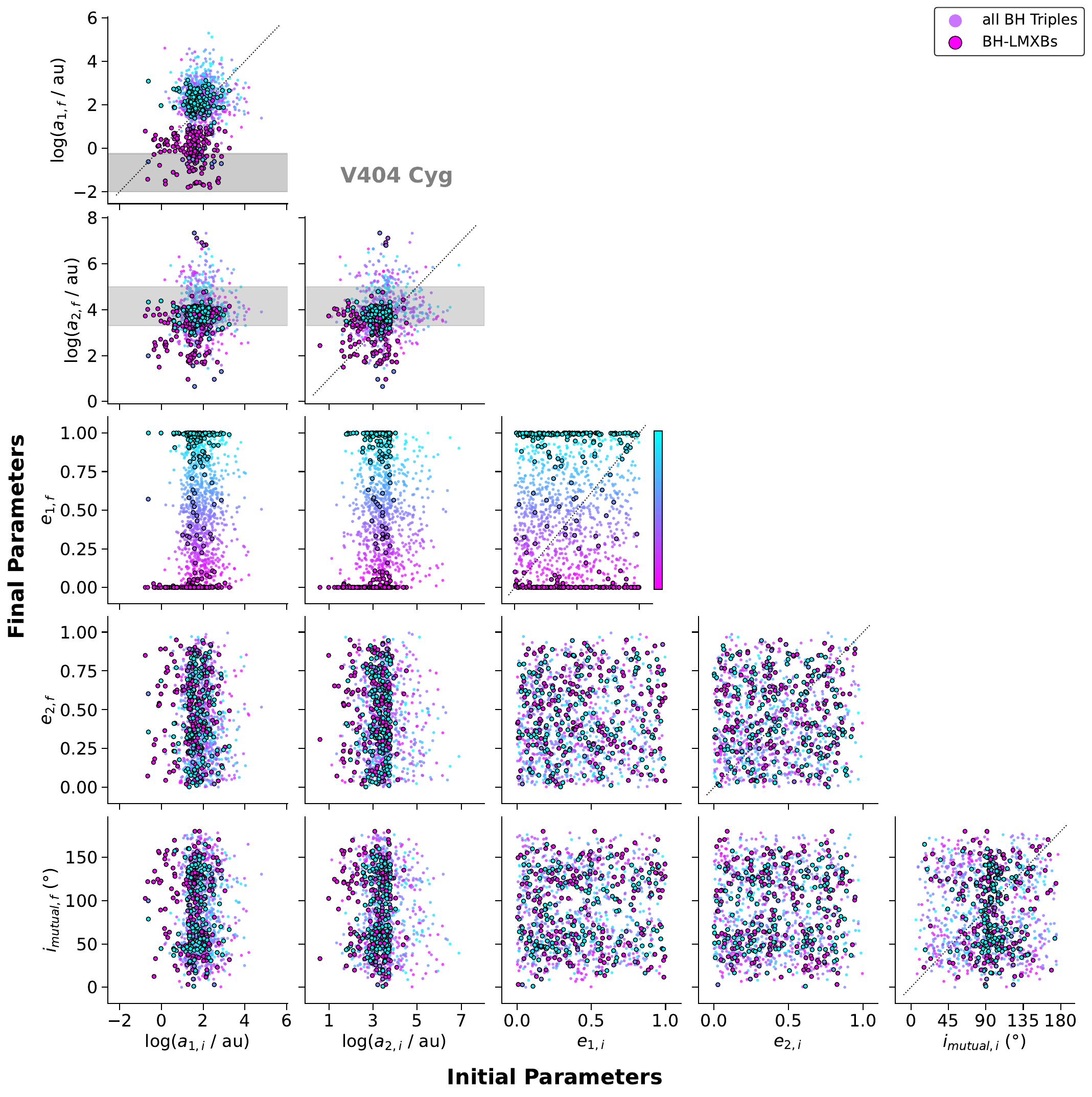}
    \caption{Corner plot comparing the initial (subscript `i`)to the final orbital parameters (subscript `f`). We plot the distributions for the initial/final inner semi-major axis ($a_1$), outer semi-major axis ($a_2$), inner eccentricity ($e_1$), outer eccentricity ($e_2$), and mutual inclination ($i_{mutual}$). We show all of the triples with a BH primary in the inner binary and highlight those that become BH-LMXBs in black. The color of each point corresponds to the final $e_1$. In the diagonal plots, we show the $1:1$ line. ) }\label{fig:corner_if} 
\end{figure}

Among our entire population synthesis of BH triples, approximately $10\%$ of the detached BH inner binaries began transferring mass while the secondary was an MS or RG star, rendering the system a BH-LMXBs. Among these mass transferring systems, $13\%~(N=42/375)$ have $a_1 <0.5$ and $a_2>1000$, giving them similar orbital configurations to the V404 Cygni hierarchical triple. One of our goals from the population of triples is to examine the relationship between the initial and final orbital parameters, focusing on systems that form BH-LMXBs and V404 Cygni-like triples. In Figure \ref{fig:corner_if}, we display a corner plot that includes the initial (horizontal) and final (vertical) orbital parameters of the triples, which include $a_1$, $a_2$, $e_1$, $e_2$, and $i_{mutual}$. We outline the BH-LMXBs in black and color each point by their final eccentricity, shown in the third row. Since these are the orbital parameters at the onset of initial mass transfer, we again see two distinct populations based on their final $a_1$ values. The systems with $\log_{10}(a_{1,f}~/~au)>\gsim1$ are mostly all highly eccentric, while those with $\log_{10}(a_{1,f}~/~au)\lsim1$ are nearly all circular. The latter population likely also experienced moderately high eccentricities ($e_1>0.9$), but became tidally circularized before mass transfer began. This population is most often associated with evolved secondaries in the inner binary since the secondary's expansion most often initiated Roche Lobe crossing (see Figure \ref{fig:Cyg404_a1_e1_a2}).

\begin{figure}[h]
    \centering
    \includegraphics[width=1.0
    \textwidth]{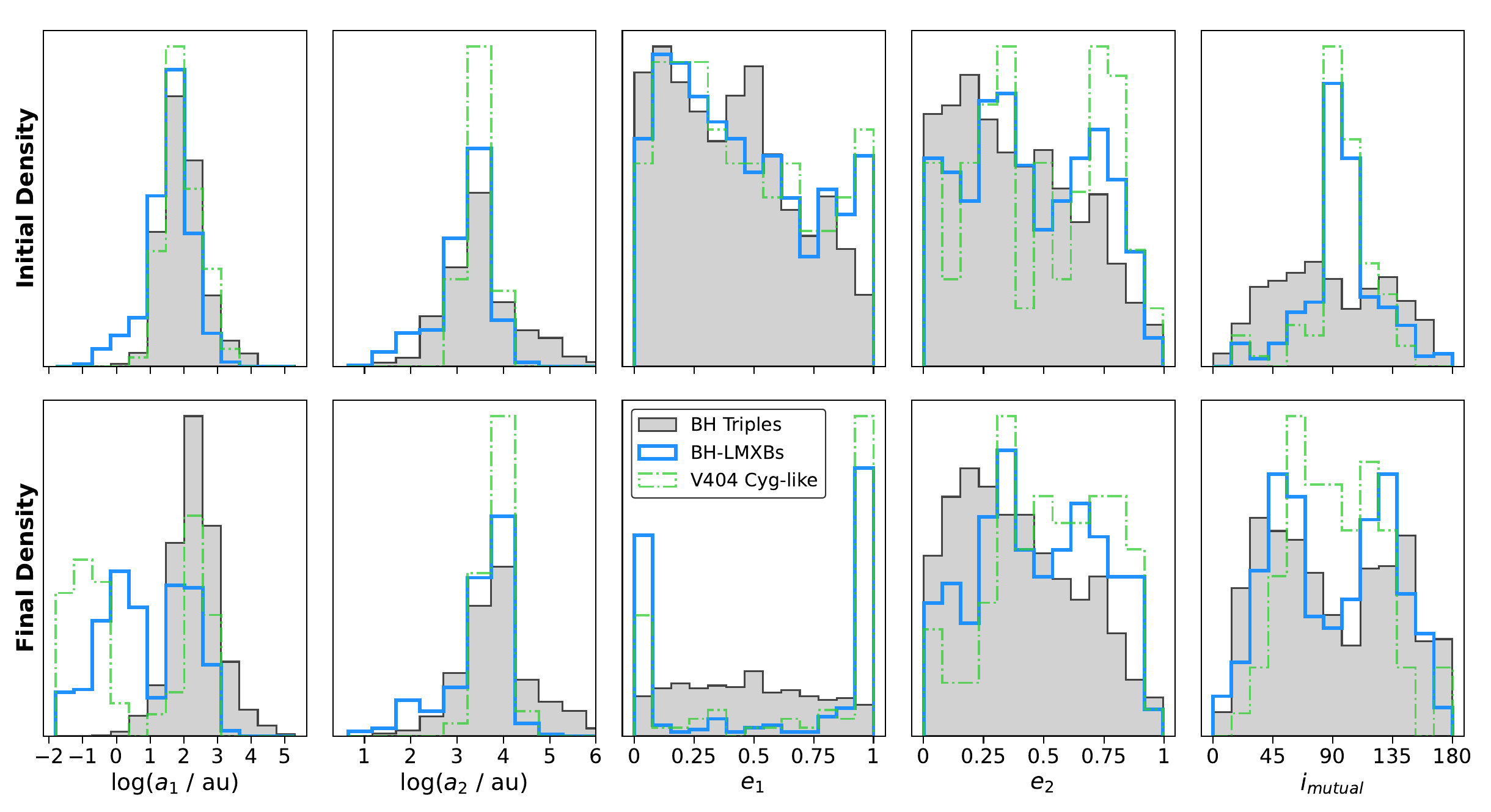}
    \caption{Histograms of initial (top) and final (bottom) probability density distributions of triples with a black hole in the inner binary. We include the triples that have detached inner binaries (gray), BH-LMXB inner binaries (blue), and BH-LMXB inner binaries with similar orbits to V404 Cygni (green dashed). All of the systems here did not experience a early mass transfer before BH formation. The parameters here are identical to those in Figure \ref{fig:corner_if}. \label{fig:404Cig_hist_if}}
\end{figure}

Another notable feature of Figure \ref{fig:corner_if} is the mutual inclination distribution for the BH-LMXBs compared to the general population of BH triples. The initial mutual inclination for BH-LMXB forming triples has a strong preference for $i_{mutual} \approx 90^{\circ}$ (also see Figure \ref{fig:404Cig_hist_if}, last column). At $90^{\circ}$, EKL-induced perturbations to the inner binary are enhanced, both to the quadrupole and octupole level \citep[e.g.,][]{Naoz13EKL}.
The bimodality in the final mutual inclinations, $i_{mutual, f}$, is observed in both populations (Figure \ref{fig:404Cig_hist_if}, last column), but is more pronounced in the BH-LMXB population.
We show the histograms from the aforementioned orbital parameters in the initial-final histograms shown in Figure \ref{fig:404Cig_hist_if}. 

Here, the gray distribution shows all BH Triples, the blue shows BH-LMXBs, and the green dashed histogram shows BH-LMXBs with triple orbital configurations similar to V404 Cygni ($a_1<0.5$~au and $a_1>2000$~au). Similar to the previous Figure, all of the data shown here is at the onset of mass transfer in from the simulations. This means that all BH+MS and BH+RG mass transferring systems are categorized as BH-LMXBs because their periastron decreases after tidal evolution. The criteria for being `V404 Cygni-like' is for the post-tides orbit, so some  V404 Cygni-like triples have large $a_1$ values before tides (e.g., column 1 in Figure \ref{fig:404Cig_hist_if}). These systems have correspondingly high eccentricities caused by secular dynamics from the tertiary, making their periastron distances near the Roche limit of the binary. Within in each column, the histograms have the same bins.

From this plot, we can make several predictions about the initial orbital structure of triples that produce BH-LMXBs and the subset of which shape into V404 Cygni-like systems. Firstly, in $a_1$ distribution, all BH-LMXBs have preferred initial separations around $a_1\sim60$~au ($\log_{10}(a_1~/~au)\sim1.8$), slightly smaller than the total population of detached BH Triples. The final $a_1$ distribution is bimodal for the BH-LMXBs, while it is roughly Gaussian for the general population. The bimodality reflects the two formation channels, where the closest inner binaries often circularized prior to mass transfer and only began transferring mass when the secondary evolved into Roche Lobe filling. In contrast, the widest systems began mass transfer on a high eccentricity excursion, as noted by the final $e_1$ distribution (column 3). In fact, most V404-like systems, and most BH-LMXBs in general, formed from this channel, where extreme eccentricities led to close pericenter passages, which either initiated mass transfer or rapid tidal locking. 

For the outer orbits semi-major axis ($a_2$), the BH-LMXB forming triples have closer separations to the inner binary, which cause stronger eccentricity excitations. The V404 Cygni-like BH-LMXBs initially have a sharp peak around $a_2\sim3000$~au, which is also the most common outer separation for all BH triple. The higher concentration of outer separations in this region also reflects that the surviving inner binaries were wide, requiring proportionately wide outer binaries for long-term stability. The initial inner eccentricities for all BH-LMXBs have a slight preference for larger values, which would allow for even small oscillations to ignite stellar interactions. For $e_2$, BH-LMXBs are most often formed with outer eccentricities around $0.3$ and $0.8$, especially for triple that become V404 Cygni-like. Since the angular momentum of the outer orbit is larger than the inner orbit's, these eccentricities are relatively conserved throughout the secular evolution, so the final $e_2$ values are also moderate. From the green curves in this column, we predict that V404 Cygni's outer tertiary began with $e_2\sim0.3$ or $e_2\sim0.8$ and today has a most-probable eccentricity in the range $e_2=0.3$ -- $0.9$. 
Relative to our entire population of LMXBs, V404 Cygni has a very common orbital architecture in nearly all regards. Many BH-LMXBs may harbor distant companions that are not yet resolved or have since been unbound. Although the initial conditions vary slightly from those in \citet{NaozLMXB}, the predictions made here are largely consistent with their study. The narrower range of masses studied applies more directly to V404 Cygni, and therefore, the time of LMXB formation changes slightly. Also, \citet{NaozLMXB} use {\tt SSE}, which can give qualitatively different outcomes that {\tt POSYDON} (Appendix \ref{app:posydon_vs_sse}).
\clearpage

\bibliography{references}
\end{document}